**Anisotropy links cell shapes to tissue flow during convergent extension**

Xun Wang[1,4], Matthias Merkel[2,3,4], Leo B. Sutter [2,4], Gonca Erdemci-Tandogan[2], M. Lisa Manning[2], Karen E. Kasza[1,*]

[1] Department of Mechanical Engineering, Columbia University, New York, New York 10027, USA

[2] Department of Physics and BioInspired Institute, Syracuse University, Syracuse, New York, 13244, USA

[3] Aix Marseille Univ, Université de Toulon, CNRS, CPT, Turing Center for Living Systems, Marseille, France

[4] These authors contributed equally: Xun Wang, Matthias Merkel, Leo B. Sutter

* To whom correspondence should be addressed. Email: karen.kasza@columbia.edu

Keywords: epithelia, morphogenesis, vertex models, *Drosophila*

**<u>Abstract</u>**

Within developing embryos, tissues flow and reorganize dramatically on timescales as short as minutes. This includes epithelial tissues, which often narrow and elongate in convergent extension movements due to anisotropies in external forces or in internal cell-generated forces. However, the mechanisms that allow or prevent tissue reorganization, especially in the presence of strongly anisotropic forces, remain unclear. We study this question in the converging and extending *Drosophila* germband epithelium, which displays planar polarized myosin II and experiences anisotropic forces from neighboring tissues, and we show that in contrast to isotropic tissues, cell shape alone is not sufficient to predict the onset of rapid cell rearrangement. From theoretical considerations and vertex model simulations, we predict that in anisotropic tissues two experimentally accessible metrics of cell patterns—the cell shape index and a cell alignment index—are required to determine whether an anisotropic tissue is in a solid-like or fluid-like state. We show that changes in cell shape and alignment over time in the *Drosophila* germband predict the onset of rapid cell rearrangement in both wild-type and *snail twist* mutant embryos, where our theoretical prediction is further improved when we also account for cell packing disorder. These findings suggest that convergent extension is associated with a transition to more fluid-like tissue behavior, which may help accommodate tissue shape changes during rapid developmental events.

**<u>Significance</u>**

Cells and tissues dramatically change shape to form functional tissues and organs during embryonic development. It is not well understood how mechanical and biological factors influence whether a developing tissue flows like a fluid or instead resists shape changes like a solid. Combining experimental studies in the fruit fly embryo with modeling approaches, we show that the shapes and alignment of cells within tissues can help to elucidate and predict how tissues change shape during development and how defects in these processes can result in abnormalities in embryo shape. Because many genes and cell behaviors are shared between fruit flies and humans, these results may reveal fundamental mechanisms underlying human development.

## Introduction

The ability of tissues to physically change shape and move is essential to fundamental morphogenetic processes that produce the diverse shapes and structures of tissues in multicellular organisms during development (1, 2). Developing tissues are composed of cells that can dynamically change their behavior and actively generate forces to influence tissue reorganization and movement (3–8). Remarkably, tissues dramatically deform and flow on timescales as short as minutes or as long as days (6). Recent studies highlight that tissue movements within developing embryos can be linked with the tissue fluidity (8–11), and computational models assuming predominantly fluid-like tissue behavior predict aspects of tissue movements (12, 13). Fluid-like tissues accommodate tissue flow and remodeling, while solid-like tissues resist flow. Yet, the mechanisms underlying the mechanical behavior of developing tissues remain poorly understood, in part due to the challenges of sophisticated mechanical measurements inside embryos and the lack of unifying theoretical frameworks for the mechanics of multicellular tissues (6, 7, 14).

Epithelial tissue sheets play pivotal roles in physically shaping the embryos of many organisms (2), often through convergent extension movements that narrow and elongate tissues. Convergent-extension is highly conserved and used in elongating tissues, tubular organs, and overall body shapes (15). Convergent extension movements require anisotropies in either external forces that deform the tissue or asymmetries in cell behaviors that internally drive tissue shape change. Indeed, an essential feature of many epithelia *in vivo* is anisotropy in the plane of the tissue sheet, a property known as *planar polarity*, which is associated with the asymmetric localization of key molecules inside cells (16–19). For example, during *Drosophila* body axis elongation, the force-generating motor protein myosin II is specifically enriched at cell edges in the epithelial germband tissue that are oriented perpendicular to the head-to-tail body axis (20, 21) (Fig. 1*A*). Planar polarized myosin is required for cell rearrangements that converge and extend the tissue to rapidly elongate the body and is thought to produce anisotropic tensions in the tissue (12, 13, 21–25). In addition, the *Drosophila* germband experiences external forces from neighboring tissues, including the mesoderm and endoderm, which have been linked to cell shape changes in the germband during convergent extension (26–29) (Fig. 1*A*). Despite being fundamental to epithelial tissue behavior *in vivo*, it is unclear how such anisotropies arising from internal myosin planar polarity and external forces influence epithelial tissue mechanical behavior, particularly whether the tissue behaves more like a fluid or a solid.

Vertex models have proven a useful framework for theoretically studying the mechanical behavior of confluent epithelial tissues (30, 31), including the packings of cells in tissues (32–34) and the dynamics of remodeling tissues (23, 32, 35–37). Recent studies of the energy barriers to cell rearrangement in isotropic vertex models, which assume no anisotropy in either internal tensions at cell-cell contacts or in external forces, have revealed a transition from solid to fluid behavior, which depends on whether large or small contacts are favored between neighboring cells. The transition is indicated by a single parameter describing cell shape, $\bar{p}$, which is the average value in the tissue of cell perimeter divided by the square root of cell area (38–40). When cells prefer smaller contacts with neighbors, $\bar{p}$ is small and the tissue is solid-like. Above a critical value of $\bar{p}$, $p_o^*$, the tissue becomes fluid-like. The isotropic vertex model successfully predicts that cell shapes identify the transition from fluid-like to solid-like behavior in cultured primary bronchial epithelial tissues; initial modeling work suggested the critical cell shape $p_o^*$ is close to 3.81 (39), in good agreement with the experiments (41). Such a simple way to infer tissue behavior from static images is appealing, particularly for tissues that are inaccessible to mechanical measurements or live imaging.

However, subsequent work has shown that the precise value of $p_o^*$ depends on specific features of the cell packing, such as the number of manyfold coordinated vertices (42) or the distribution of neighbor numbers in the packing (43–45), though the latter feature has never been studied systematically. In addition, these previous vertex model studies did not account for effects of anisotropy, potentially limiting their use in the study of converging and extending tissues.

Here, we combine confocal imaging and quantitative image analysis with a vertex model of anisotropic tissues to study epithelial convergent extension during *Drosophila* body axis elongation. We show that cell shape alone is not sufficient to predict the onset of rapid cell rearrangement during convergent extension in the *Drosophila* germband, which exhibits anisotropies arising from internal forces from planar polarized myosin and external forces from neighboring tissue movements. Instead, we show that for anisotropic tissues, such as the *Drosophila* germband, anisotropy shifts the predicted transition between solid-like and fluid-like behavior and so must be taken into account, which can be achieved by considering both cell shape and cell alignment in the tissue. We find that the onset of cell rearrangement and tissue flow during convergent extension in wild-type and mutant *Drosophila* embryos is more accurately described by a combination of cell shape and alignment than by cell shape alone. Moreover, we use experimentally accessible features of cell neighbor relationships to quantify cell packing disorder and pinpoint $p_o^*$, which further improves our predictions. These findings suggest that convergent extension is associated with a transition from solid-like to more fluid-like tissue behavior, which may help to accommodate dramatic epithelial tissue shape changes during rapid axis elongation.

## Results

**Cell shape alone is not sufficient to predict the onset of rapid cell rearrangement in the *Drosophila* germband epithelium.** To explore the mechanical behavior of a converging and extending epithelial tissue *in vivo*, we investigated the *Drosophila* germband, a well-studied tissue that has internal anisotropies arising from planar polarized myosin (20–25, 46) and also experiences external forces from neighboring developmental processes that stretch the tissue (26, 27). The germband rapidly extends along the anterior-posterior (AP) axis while narrowing along the dorsal-ventral (DV) axis (Fig. 1*A*), roughly doubling the length of the head-to-tail body axis in just 30 min (47) (Fig. 1*C*). Convergent extension in the *Drosophila* germband is driven by a combination of cell rearrangements and cell shape changes (Fig. 1*B,C*). The dominant contribution is from cell rearrangement (21, 22, 28, 47), which requires a planar polarized pattern of myosin localization across the tissue (20, 21) that is thought to be the driving force for rearrangement (21, 23, 24, 46). Cell stretching along the AP axis also contributes to tissue elongation and coincides with movements of neighboring tissues (26–28, 48, 49), indicating that external forces play an important role in tissue behavior. Despite significant study of this tissue, a comprehensive framework for understanding its mechanical behavior is lacking, in part because direct mechanical measurements inside the *Drosophila* embryo, and more generally for epithelial tissues *in vivo*, continue to be a challenge (50–52).

To gain insight into the origins of mechanical behavior in the *Drosophila* germband epithelium, we first tested the theoretical prediction of the vertex model that cell shapes can be linked to tissue mechanics. In the isotropic vertex model, tissue mechanical behavior is reflected in a single parameter, the average cell shape index $\bar{p}$ (38–41). To quantify cell shapes in the *Drosophila* germband, we used confocal time-lapse imaging of embryos with fluorescently-tagged cell membranes (53) and segmented the resulting time-lapse movies (28) (Fig. 2*A*, *SI Appendix* Fig. S1). Prior to the onset of tissue elongation, individual cells take on roughly isotropic shapes and become more elongated over time (Fig. 2*A,B*), consistent with previous observations (26–28, 54). Ten minutes prior to tissue elongation, the cell shape index $\bar{p}$ averaged over 8 wild-type embryos is just above 3.81. Eight minutes before the onset of tissue elongation, $\bar{p}$ starts to increase before reaching a steady value of 3.98 about 20 min after the onset of tissue elongation (Fig. 2*B*). The average cell shape index prior to tissue elongation, $\bar{p} = 3.81$ (dashed line, Fig. 2*B*), is close to the value associated with isotropic solid-like tissues in previous work (38–40), suggesting that the tissue may be solid-like prior to elongation.

We next asked how these cell shapes vary among the individual embryos and correlate with tissue mechanical behavior. As an experimentally accessible read-out of tissue fluidity, we used the instantaneous rate of cell rearrangements occurring within the germband tissue (Fig. 1*C*), where higher rearrangement rates are associated with more fluid-like behavior and/or larger driving forces. Plotting instantaneous cell rearrangement rate versus $\bar{p}$ at each time point from movies of individual wild-type *Drosophila* embryos, we find that the onset of rapid cell rearrangement occurs at different values of $\bar{p}$ for each embryo, ranging from 3.83 to 3.90 for a cutoff rearrangement rate per cell of 0.02 min$^{-1}$ (Fig. 2*C*). We verified that we observe a similar variation in the values of $\bar{p}$ for different cutoff values (*SI Appendix* Fig. S2). This suggests that in the germband epithelium, comparing the cell shape index $\bar{p}$ to a fixed critical value (e.g. 3.81) is not sufficient to predict tissue behavior.

**Cellular packing disorder is not sufficient to predict the onset of rapid cell rearrangement in the germband.** Recent vertex model simulations suggest that $\bar{p} = 3.81$ is often insufficient to separate solid from fluid tissue behavior, as the precise location of the solid-fluid transition depends on how exactly cells are packed in the tissue (42–45). A hexagonal packing has no packing disorder, while each cell with neighbor number different from six increases the packing disorder in the tissue. In the modeling literature, this disorder is typically generated either by allowing manyfold coordinated vertices (i.e. vertices at which more than three cells meet) or using simulation preparation protocols that create cell neighbor numbers other than six. Including manyfold vertices in simulations is natural, as they are observed in the germband epithelium (54), and are often formed during cell rearrangements involving four or more cells (21, 22). Moreover, recent theoretical work has predicted how the presence of manyfold vertices increases the critical shape index (42).

We wondered whether cell packing disorder quantified by the vertex coordination number $z$ could explain the observed embryo-to-embryo variability in $\bar{p}$ at the transition point in wild-type embryos (Fig. 2*C*, *SI Appendix* Fig. S2). To test this idea, we plotted $\bar{p}$ versus $z$ at each time point and color-coded the data based on the instantaneous cell rearrangement rate, pooling the data from all wild-type embryos (Fig. 2*F*). To isolate the changes in mechanical behavior of the germband during convergent extension from later developmental events, we focus on times $t \leq 20$ min after the onset of tissue elongation, well before cell divisions begin in the germband. If vertex coordination were sufficient to explain the germband behavior, then the theoretically determined line (dashed line) should separate regions with low cell rearrangement rate (blue symbols) from regions with high cell rearrangement rate (red, orange, and yellow symbols) (Fig. 2*F*). However, this is not the case, indicating that the prediction from Ref. (42) alone is not sufficient to account for the germband behavior during this stage.

Next, we asked if other aspects of packing disorder could affect tissue fluidity. Even without manyfold vertices, it is possible to generate packings *in silico* with differences in packing disorder just by altering the preparation protocol. Since this has not been systematically studied, we performed a large number of vertex model simulations where we varied the packing disorder (*SI Appendix* Fig. S3*A,B*). In our simulations, the transition point is well predicted by the fraction of pentagonal cells, i.e. cells that have exactly five neighbors, with a linear dependence (Fig. 2*D*, *SI Appendix* Fig. 3*C,D*). In particular, without any pentagonal cells, we recover the previously predicted transition point of $\approx 3.72$ for tissues consisting only of hexagonal cells (33, 38). In comparison, the previously reported value of 3.81 corresponds to a fraction of $\approx 15\%$ pentagonal cells (Fig. 2*D*). While additional aspects of cell packing likely affect the transition, these results suggest that the fraction of pentagonal cells may also be a good predictor for the transition point in isotropic tissues.

To test whether this second measure of packing disorder could explain the variability in $\bar{p}$ at the transition in wild-type embryos, we plotted $\bar{p}$ versus the fraction of pentagonal cells at each time point, again color-coding the data based on the instantaneous cell rearrangement rate and pooling data from all wild-type embryos (Fig. 2*E*, *SI Appendix* Fig. S2). We find that the packing disorder quantified by the fraction of pentagonal cells is also insufficient to explain the onset of cell rearrangements.

Our results suggest that two measures of packing disorder, the vertex coordination number and fraction of pentagonal cells, have at least partially independent effects on the isotropic vertex model transition point. However, neither of them is sufficient to understand the transition to high cell rearrangement rates in the *Drosophila* germband.

**Theoretical considerations and vertex model simulations predict a shift of the solid-fluid transition in anisotropic tissues.** To study whether anisotropies in the germband could affect the relation between the cell shape index and cell rearrangement rate, we used vertex model simulations to test how tissue anisotropy, introduced into the model in different ways, affects tissue fluidity.

First, we introduced anisotropy by applying an external deformation, mimicking the effects of forces exerted by neighboring morphogenetic processes, and then studied force-balanced states of the model tissue (Fig. 3*A*). As a metric for tissue stiffness, we measured the shear modulus of the model tissue, which describes with how much force a tissue resists changes in shape. A vanishing shear modulus corresponds to fluid behavior, where the tissue flows and cells rearrange in response to any driving force, whereas a positive shear modulus indicates solid behavior, where the tissue does not flow so long as the driving force is not too large. We then analyzed how the shear modulus correlates with $\bar{p}$ for different amounts of global tissue deformation, quantified by the strain $\varepsilon$ (Fig. 3*B*). For small strain, we recover the behavior of the isotropic vertex model. The shear modulus is finite when $\bar{p}$ is small and vanishes above a critical cell shape index, which is $p_o^* = 3.94$ for our simulations (Fig. 3*B*, blue symbols). For larger strains, we find that the critical value of the shape index at the transition between solid-like and fluid-like behavior generally increases with the amount of strain (Fig. 3*B*). Indeed, $\bar{p}$ for cells in a deformed, solid tissue can be higher than for cells in an undeformed, fluid tissue. This suggests that anisotropy affects the critical shape index at which the tissue transitions between solid and fluid behavior.

Some of us recently developed a theoretical understanding for a shift in the critical shape index when deforming a vertex model tissue (45). In the limit of small deformations by some strain ε and without cell rearrangements, the critical value of $\bar{p}$ increases from $p_o^*$ to $p_o^* + b\varepsilon^2$, where $b$ is a constant prefactor. To compare this formula to the vertex model simulations (Fig. 3*A,B*), we need to take into account that cell rearrangements occur in our simulations. Removing their contribution from the overall tissue strain ε leaves us with a parameter $Q$ (Fig. 3*C*) (*SI Appendix*, SI Materials and Methods), which can be quantified using a triangulation of the tissue created from the positions of cell centers (*SI Appendix*, SI Materials and Methods) (55, 56). We term $Q$ a "shape alignment index", as $Q$ is non-zero only when the long axes of cells are aligned. We emphasize that, unlike the nematic order parameter for liquid crystals, the cell alignment parameter $Q$ is additionally modulated by the degree of cell shape anisotropy; tissues with the same degree of cell alignment but more elongated cells have a higher $Q$ (Fig. 3*C*). In other words, $Q$ can be regarded as a measure for tissue anisotropy. After accounting for cell rearrangements, we expect the transition point in anisotropic tissues to shift from the isotropic value $p_o^*$ to (*SI Appendix,* SI Materials and Methods):

$$\bar{p}_{crit} = p_o^* + 4bQ^2 . \qquad (1)$$

Indeed, comparing this equation to vertex model simulations yields a good fit with the simulation results (Fig. 3*D*, solid line), with fit parameters $p_o^* = 3.94$ and $b = 0.43$ (*SI Appendix* Fig. S4). We confirmed that cell area variation did not significantly affect these findings (*SI Appendix* Fig. S5). In principle, we expect both the transition point $p_o^*$ and the precise value of $b$ to depend on the packing disorder, but our best-fit value for $b$ is consistent with previously published results (45). Therefore, we used $b = 0.43$ for the remainder of this study. Hence, for external deformation, the solid-fluid transition point in the vertex model increases quadratically with tissue anisotropy $Q$.

We also tested how the model predictions change when we introduce anisotropy generated by internal forces into the vertex model. We modeled myosin planar polarity as increased tensions on "vertical" cell-cell contacts (Fig. 3*E*, *SI Appendix* Fig. S6) (23), and focus again on stationary, force-balanced states. We investigated simulations of model tissues with internal forces, both with (Fig. 3*E*) and without (Fig. 3*E*, *inset*) externally applied deformation. We find that in both cases solid states exist for larger cell shape indices than the isotropic $p_o^* = 3.94$, and our results are again consistent with the fit from Fig. 3*D* (solid lines in Fig. 3*E* and *inset*). With finite anisotropic internal tensions only, we obtain states in the fluid regime that do not reach a force-balanced state (detailed discussion in *SI Appendix*), and this explains the white region devoid of stable states in the upper middle region of Fig. 3*E, inset*. Taken together, these findings demonstrate that a combination of cell shape $\bar{p}$ and cell shape alignment $Q$ in the vertex model indicates whether an anisotropic tissue is in a solid-like or fluid-like state, regardless of the underlying origin of anisotropy.

**Cell shape and cell shape alignment together indicate the onset of cell rearrangement during *Drosophila* axis elongation.** We returned to our experiments to test whether a combination of $\bar{p}$ and $Q$ would be a better predictor for the behavior of the *Drosophila* germband during convergent extension. We quantified alignment $Q$ using the triangle method (Fig. 4*A*) and found that prior to the onset of tissue elongation, which begins at $t = 0$ min, alignment is not very high (Fig. 4B). $Q$ begins to increase just prior to elongation, peaking at $t = 1$ min (Fig. 4B, *SI Appendix* Fig. S7), which is consistent with observations using other cell pattern metrics (23, 26, 28, 29). This peak in $Q$ corresponds to stretching of cells along the dorsal-ventral axis, perpendicular to the axis of germband extension, and coincides with the time period during which the presumptive mesoderm is invaginating (29, 53). $Q$ relaxes back to low levels during axis elongation (Fig. 4B). Plotting $\bar{p}$ vs $Q$ at each time point from movies of individual wild-type embryos reveals common features, despite embryo-to-embryo variability (Fig. 4C, *inset*). Initially, we see a concomitant increase of $\bar{p}$ and $Q$ prior to the onset of convergent extension. Above $\bar{p} = 3.87$, $Q$ decreases drastically as $\bar{p}$ continues to increase, indicating that further increases in $\bar{p}$ are associated with randomly oriented cell shapes (cf. Fig. 3C). Thus, cell shapes in the germband are transiently aligned around the onset of convergent extension.

We next asked whether this temporary increase in alignment could help resolve the seeming contradiction between the measured cell shapes and cell rearrangement rates. To this end, we investigated how $\bar{p}$ and $Q$ correlate with the instantaneous rate of cell rearrangements occurring within the germband, with higher rearrangement rates associated with more fluid-like behavior and/or larger active driving forces (Fig. 4C). The anisotropic vertex model predicts that the solid or fluid behavior of the tissue should depend on both $\bar{p}$ and $Q$ according to Eq. (1), with only two adjustable parameters, $p_o^*$ and $b$. We fit Eq. (1) to our experimental data by minimizing a quality of fit measure defined as the number of experimental data points on the wrong side of the theoretical transition line, and for simplicity vary only $p_o^*$ while keeping the theoretically determined value for $b$. Varying the value $b$ leads at most to a slight improvement of our fit (*SI Appendix* Fig. S8). To differentiate between solid-like and fluid-like tissue behavior in the experimental data, we need to choose a cutoff value for the cell rearrangement rate. Choosing a cutoff of 0.02 min$^{-1}$ per cell yields a best fit with $p_o^* = 3.83$ (solid line, Fig. 4C). To confirm that our prediction of a quadratic dependence on $Q$ is supported by the data, we also identify the best fit to a null hypothesis of a $Q$-independent transition point (horizontal dashed line, Fig. 4C). Using our quality of fit measure, we find that the $Q$-dependent fit is always better, independent of the chosen cell rearrangement rate cutoff (*SI Appendix* Fig. S8).

Comparing the trajectories of individual embryos (Fig. 4C, *inset*) to the predicted transition in the anisotropic vertex model (Fig. 4C), we see that during early times, when $\bar{p}$ and $Q$ are both increasing, the tissue stays within the predicted solid-like regime. The subsequent rapid decrease in $Q$ brings embryos closer to the transition line. As $\bar{p}$ further increases, individual embryos cross this transition line, which coincides with increased rates of cell rearrangement, at different points $(Q, \bar{p})$. Thus, compared to the isotropic model, the anisotropic vertex model better describes the onset of rapid cell rearrangement and tissue flow during convergent extension with two metrics of cell patterns, $\bar{p}$ and $Q$, that are both easy to access experimentally.

**Accounting for cell shape alignment and cell packing disorder allows for a parameter-free prediction of tissue behavior.** While the above results confirm that tissue anisotropy must be taken into account to predict the onset of rapid cell rearrangement, the theoretical prediction in Fig. 4C still required a fit parameter $p_o^*$. Theoretical results suggest that this fit parameter, which is the *isotropic* transition point in the absence of anisotropic forces, should depend systematically on cell packing disorder quantified by vertex coordination (42) and fraction of pentagonal cells (Fig. 2D).

Therefore, we analyzed the $\bar{p}$ and $Q$ data for each embryo individually, by fitting them to Eq. (1) with $b$=0.43 where we again use $p_o^*$ as the only fit parameter (Fig. 4D, *inset*). We compared the $p_o^*$ obtained for each embryo (purple point, Fig. 4D *inset*) to the average vertex coordination number in the tissue at the time of the transition (green point, Fig. 4D *inset*) and found a clear correlation (dashed line, Fig 4F), which fits well with the previous theoretical prediction (42), with no fit parameters.

Combining this previous theoretical prediction of the effects of vertex coordination on the solid-fluid transition in isotropic tissues with our prediction for how cell shape alignment shifts this transition in anisotropic tissues in Eq. (1) generates the following parameter-free prediction of the critical shape index for tissue fluidity:

$$\bar{p}_{crit} = 3.818 + (z-3)/B + 4bQ^2, \qquad (2)$$

where $z$ is the measured average vertex coordination number, and the other parameters are universally determined *a priori* from vertex model simulations: $B = 3.85$ (42), and $b = 0.43$. To test this prediction, we plot the cell shape index corrected by the vertex coordination number, $\bar{p}_{corr} = \bar{p} - (z-3)/B$, versus cell shape alignment $Q$ in the germband of wild-type embryos, and compare it to the theoretical curve given by $\bar{p}_{corr} = 3.818 + 4bQ^2$ (solid line, Fig. 4D). Remarkably, this parameter-free prediction describes our experimental data well. We compared the quality of fit to alternative parameter-free predictions and found that Eq. (2) consistently provides the best prediction for a wide range of cell rearrangement rate cutoffs (*SI Appendix* Fig. S8).

Some embryos deviate from the theoretical prediction from Ref. (42) (Fig. 4F), suggesting that perhaps alternate features of packing disorder may play an important role in those embryos. Thus we also compared $p_o^*$ obtained from the individual-embryo fits to the respective fraction of pentagons at the time of the transition, and found a strikingly clear correlation well described by a linear relation (Fig 4E, dashed line is a linear fit). This relationship quantitatively differs from what we extracted from our vertex model simulations (Fig. 2D), indicating again that other aspects of packing disorder may also play a role. Nevertheless, using this linear fit to correct the shape index for each data point by the fraction of pentagonal cells, we obtain an improved prediction of our data (compare Fig. 4D to *SI Appendix* Fig. S9) at the expense of requiring two fit parameters.

Taken together, these results show that we can quantitatively predict the behavior of the germband tissue in wild-type embryos, with no fit parameters using Eq. (2), from an image of cell patterns in the tissue. To do so, we needed to quantify three observables: cell shapes, cell alignment, and cell packing disorder. We found that vertex coordination and the fraction of pentagonal cells are both good proxies for packing disorder, in vertex model simulations and the germband.

**Cell shape, alignment, and tissue behavior in *snail twist* and *bnt* mutant embryos**. Since the *Drosophila* germband experiences both internal forces due to myosin planar polarity and external forces from neighboring tissues, we wondered whether our theoretical predictions hold when altering the nature of the forces in the germband. To dissect the effects of internal and external sources of tissue anisotropy, we studied cell patterns in *snail twist* mutant embryos, which lack genes required for invagination of the presumptive mesoderm (57), and in *bcd nos tsl* (*bnt*) mutant embryos, which lack patterning genes required for planar polarized patterns of myosin localization and axis elongation (22, 47).

First, we analyzed cell shapes and cell shape alignment in the germband of *snail twist* mutant embryos in which the presumptive mesoderm does not invaginate. In *snail twist* embryos, we observe that the germband tissue elongates (Fig. 5C) and cell rearrangements occur (Fig. 5D), similar to prior studies (28) although at somewhat reduced rates compared to in wild-type embryos. However, in contrast to wild-type embryos, we find that the cell shape alignment $Q$ is significantly reduced between $t = -5$ min and $t = +8$ min (Fig. 5A,F), similar to previous reports of other metrics for cell stretching (28). The cell shape index $\bar{p}$ is also reduced during this period (Fig. 5E). These observations are consistent with the idea that external forces from mesoderm invagination produce the transient cell shape elongation and alignment observed in wild-type embryos.

Next, we tested whether our theoretical predictions would describe tissue behavior in *snail twist* embryos, even with their significantly reduced cell alignment. We found that the onset of rapid cell rearrangement in *snail twist* embryos is also well predicted by Eq. (2) (Fig. 5G). This is corroborated by comparing the parameters $p_o^*$ of the individual *snail twist* embryo fits to the vertex coordination number at the transition (Fig. 5G, inset), which is close to the previous theoretical prediction (dashed line) (42). Hence, our prediction also holds in embryos with reduced cell shape alignment $Q$, where the transition to rapid cell rearrangement occurs at a lower cell shape index $\bar{p}$ compared to in wild-type embryos (Fig. 5H).

To investigate how disrupting other forces in the germband affects tissue behavior, we studied cell patterns in *bnt* mutant embryos, which lack anterior-posterior patterning genes required for axis elongation. These mutant embryos do not display myosin planar polarity, although there is significant myosin present at the apical cortex of cells (*SI Appendix* Fig. S10). The *bnt* embryos have severe defects in tissue elongation (Fig. 5C), cell rearrangement (Fig. 5D), and endoderm invagination, but still undergo mesoderm invagination (13, 20, 22, 26, 28, 47). $\bar{p}$ displays an initial increase (Fig. 5E), concomitant with an increase in $Q$ (Fig. 5F), similar to in wild-type embryos. After $t = 1$ min, $\bar{p}$ does not increase further and takes on a steady value of 3.87 (Fig. 5E). This supports the idea that the further increase in $\bar{p}$ in wild-type embryos is due to internal anisotropies associated with myosin planar polarity or external forces associated with endoderm invagination. Interestingly, $Q$ returns more slowly to low levels in *bnt* compared to wild-type embryos (Fig. 5F), suggesting a potential role for myosin planar polarity, cell rearrangements oriented along the AP axis, or endoderm invagination in relaxing cell shape alignment along the DV axis. The *bnt* tissues do not transition to a state of rapid cell rearrangement. This is not consistent with the predictions of Eq. (2) (Fig. 5I), which predicts some fluid-like tissue states in the germband of *bnt* embryos, suggesting that either the driving forces are too small or that there are additional barriers that prevent rapid cell rearrangement in these embryos.

Taken together, these findings demonstrate that external forces associated with mesoderm invagination contribute to tissue anisotropy in the germband and that the onset of rapid cell rearrangement can be predicted from cell shape and alignment, even in the absence of forces associated with mesoderm invagination.

**<u>Discussion</u>**

In this work, we show that cell shape, cell alignment, and packing disorder can be used to understand and predict whether an anisotropic tissue flows and remodels like a fluid or instead maintains its shape like a solid. Importantly, in contrast to isotropic tissues, the mechanical behavior of the converging and extending *Drosophila* germband cannot be predicted by cell shape and packing disorder alone. Instead, we show via theoretical analysis and simulation that in anisotropic tissues three experimentally accessible metrics—the cell shape index $\bar{p}$, the cell alignment index $Q$, and packing disorder quantified by either vertex coordination or fraction of pentagonal cells—are required to determine whether an anisotropic tissue flows and remodels or not. We demonstrate that the onset of rapid cell rearrangement in wild-type *Drosophila* embryos is indeed more accurately described by a combination of these three cell pattern metrics, using an equation with no fit parameters, than by cell shape or packing disorder alone. We further tested this prediction in *snail twist* mutant embryos in which the presumptive mesoderm does not invaginate and found that our parameter-free prediction successfully predicts the onset of rapid cell rearrangement and tissue flow in this case as well. These findings suggest that convergent extension of the *Drosophila* germband might be viewed as a transition to more fluid-like behavior to help accommodate dramatic tissue flows. This raises the possibility that the properties of developing tissues might be tuned to become more fluid-like during rapid morphogenetic events.

A fluid-to-solid jamming transition has recently been reported in mesodermal tissues during zebrafish body axis elongation (8). In contrast to the zebrafish mesoderm in which the transition to more solid-like behavior is associated with an increase in cellular volume fraction (proportion of the tissue occupied by cells), the *Drosophila* germband epithelium comprises tightly packed cells and its mechanical behavior changes in the absence of any change in cell volume fraction. Future studies will be needed to explore how the properties of epithelial cells might be regulated during development to tune the mechanical behaviors of the tissues in which they reside.

The vertex model predictions of tissue behavior are independent of the underlying origin of anisotropy, and therefore can be used to predict mechanical behavior of tissues from cell shape patterns, even when external and internal stresses cannot be directly measured. Although our current simulations were not able to access some of the tissue states driven by internal stresses, we found that the cases that were accessible were fully consistent with our simulation results without internal stresses. Importantly, the average cell shape index $\bar{p}$, cell shape alignment index $Q$, and metrics for packing disorder are easy to access experimentally from snapshots of cell packings in tissues, even in systems where time-lapse live imaging of cell rearrangement and tissue flow is not possible. Thus, this approach may prove useful for studying complex tissue behaviors in a broad range of morphogenetic processes occurring in developing embryos *in vivo* or organoid systems *in vitro*.

In our analysis, we characterized the mechanical state of the germband epithelial tissue using the rate of cell rearrangement as the observable. We made this choice because direct measurements of the mechanical properties of the germband remain a significant experimental challenge (6, 7, 14). Generally, higher rates of cell rearrangement could be due to more fluid tissue properties or a stronger driving force, which is the sum of externally applied forces and internally generated mechanical stresses. Based on our Eq. (2) result, the cell shape index and alignment predict the onset of rapid cell rearrangement in the germband. While this would be consistent with the tissue becoming more fluid, it is also possible that the observed increase in cell rearrangement rate is at least in part due to an increase in the driving force while the tissue remains solid.

To parse this possibility further, it is useful to consider a solid tissue, where the tissue will flow only if it is pulled with a force above some threshold called the yield stress. If the tissue is deeply in the solid state, far from the solid-fluid transition, and the applied force is far above the yield stress, one would expect cells to acquire elongated shapes and transiently form manyfold vertices during cell rearrangements in response to the applied force. The rearrangement rate would correlate with the cell shape index, after accounting for packing disorder and alignment, which is similar to what we predict with our fluid-solid model. However, based on our vertex model simulations we would not expect to see tissue states with high shape index $\bar{p}$ and low alignment $Q$ associated with high rearrangement rates for solid tissues. Since we do observe such tissue behavior during germband extension, this suggests that the germband is more fluid-like during these periods with high cell rearrangement rates.

Of course, it could be that the tissue is a very *weak* yield stress solid, so that it becomes fluid-like under very small applied forces. This is consistent with the observations that the large majority of rearrangements are oriented along the head-to-tail body axis (21, 22, 46, 47, 58) and the time period of rapid cell rearrangement (Fig. 1C) coincides with the period of planar polarized myosin (13, 25, 46). Direct mechanical measurements of the germband have not been conducted during axis elongation, but ferrofluid droplet and magnetic bead microrheology measurements have probed the mechanical behavior of the epithelium prior to germband extension in the cellularizing embryo. These studies report that tissue behavior is predominantly elastic (solid-like) over timescales less than several minutes and suggest fluid-like behavior on the longer ~30 min timescales relevant for germband extension (51, 52). These measurements might also be consistent with a weak yield-stress solid, an interpretation that would be supported by the near absence of cell rearrangements prior to germband extension. Taken together, these observations suggest that over the time period that we describe the germband as "fluid-like", it could actually be a very weak yield-stress solid.

Though there is often little functional difference between a fluid and weak yield stress solid, the difference may be relevant for mutant *bnt* embryos, whose behavior is not well-captured by our theoretical predictions. In particular, we observe *bnt* tissues with $\bar{p}$, $Q$, and cell packing disorder that would be predicted to display fluid-like behavior but do not undergo rapid cell rearrangement. This suggests that in these embryos the driving forces are not sufficient to overcome the yield stress. One obvious explanation for this is that the germband in *bnt* embryos experiences altered forces associated with disrupted myosin planar polarity (22) and defects in endoderm invagination, which would contribute to a reduced driving force. Alternatively, additional barriers to cell rearrangement in *bnt* mutants, of the sort described in Ref. (59), could also explain this behavior.

Similarly, our vertex model does not predict the observed decrease in cell rearrangement rates after 20 minutes of axis elongation (Fig. 1C). Given the observed high values of $\bar{p}$ and low values of $Q$, our model would still predict fluid-like behavior. Just as in the *bnt* mutants, this discrepancy could be explained by a decreased driving force or additional barriers to cell rearrangement. The former explanation is supported by the observation that myosin planar polarity reaches a maximum 5-10 min after the onset of axis elongation and then decreases during the rest of the process (25, 28, 46), while the latter could potentially be explained by maturation of cell junctions or changes to adhesive interactions over the course of embryonic development (60, 61).

Consistent with the notion of additional barriers to cell rearrangement, recent work suggests that local remodeling of active junctional tension at cell-cell contacts only occurs above a critical strain threshold in cultured epithelial cells (59, 62). This is consistent with a growing body of work that points toward important roles for membrane trafficking and E-cadherin turnover in junctional remodeling during *Drosophila* epithelial morphogenesis (11, 63–65). Indeed, such a mechanism of mechanosensitive barriers to junctional remodeling and cell rearrangement can be added to standard vertex models to explain such weak yield-stress behavior (59).

Moving forward, it will be interesting to explore experimentally how the nature of internal and external forces contributes to tissue mechanics, cell rearrangement, and tissue flows in the germband and other developing epithelial tissues. Incorporating these features into more sophisticated vertex models will contribute to understanding the diverse behaviors of living tissues, and the approaches we develop here will be useful for interrogating these questions.

## Methods

Embryos were generated at 23℃ and analyzed at room temperature. Cell outlines were visualized with gap43:mCherry (53), Spider:GFP, or Resille:GFP cell membrane markers. Embryos were imaged on a Zeiss LSM880 laser scanning confocal microscope. Time-lapse movies were analyzed with SEGGA software in MATLAB (28) for quantifying cell shapes and cell rearrangement rates, PIVlab version 1.41 in MATLAB (66) for quantifying tissue elongation, and custom code for quantifying cell alignment using the triangle method (55, 56, 67). The vertex model describes an epithelial tissue as a planar tiling of $N$ cellular polygons, where the degrees of freedom are the vertex positions (33). Forces in the model are defined such that cell perimeters and areas act as effective springs with a preferred perimeter $p_0$ and a preferred area of one, which is implemented via an effective energy functional (45). Unless otherwise noted, error bars are the standard deviation. The data that support the findings of this study are included in the paper and *SI Appendix*. The custom code used in this study to extract the average triangle-based $Q$ tensor from images segmented using SEGGA (28) is available at https://github.com/mmerkel/triangles-segga. Details can be found in the *SI Appendix*, SI Materials and Methods.

## Acknowledgements

The authors thank Erik Boyle for assistance with data processing; Dene Farrell and Jennifer Zallen for the use of SEGGA, a segmentation and quantitative image analysis toolset; Adam Martin for the sqh-gap43:mCherry fly stock; and the Bloomington Drosophila Stock Center (BDSC) for fly stocks. We would like to thank an anonymous reviewer of our manuscript for suggesting that we develop a more quantitative analysis of packing disorder for our data, ultimately resulting in a significant improvement in our ability to predict tissue flow. This work was supported by the National Science Foundation CMMI 1751841 to K.E.K., DMR-1352184 and POLS-1607416 to M.L.M, and DMR-1460784 (REU) to L.B.S. M.L.M., M.M., and G.E.T. acknowledge support from Simons Grant No. 446222 and 454947, and NIH R01GM117598. K.E.K. holds a BWF Career Award at the Scientific Interface, Clare Boothe Luce Professorship, and Packard Fellowship.

**Figures and captions**

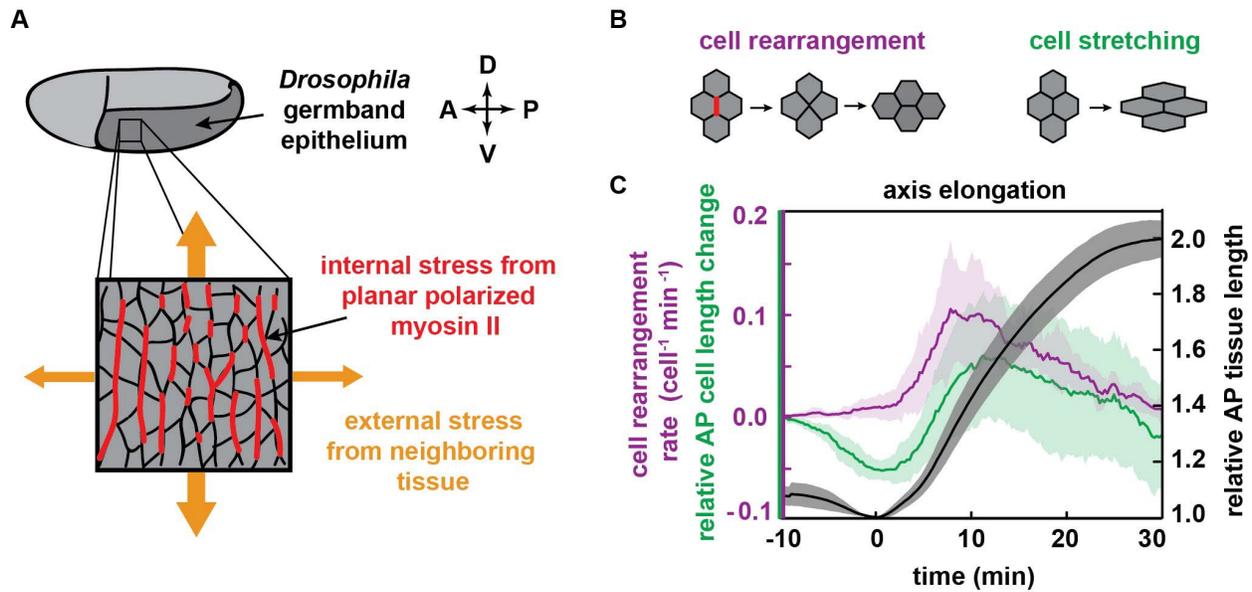

**Figure 1. Cell shapes and cell rearrangements in the converging and extending *Drosophila* germband epithelium during axis elongation.** (*A*) Schematic of *Drosophila* body axis elongation. The germband epithelium (dark gray) narrows and elongates along the head-to-tail body axis in a convergent extension movement. The tissue is anisotropic, experiencing internal stresses from planar polarized patterns of myosin II (red) within the tissue as well as external stresses (orange) due to the movements of neighboring tissue. (*B*) Schematic of oriented cell rearrangement and cell shape change. (*C*) The germband epithelium doubles in length along the head-to-tail AP axis in 30 min (black). Cell rearrangements are thought to drive tissue elongation (magenta), and cell shape changes also contribute (green). Tissue elongation begins at $t = 0$. The cell rearrangement rate includes cell neighbor changes through T1 processes and higher order rosette rearrangements. Relative cell length along the AP axis is normalized by the value at $t = -10$ min. Mean and standard deviation between embryos is plotted (*N*=8 embryos with an average of 306 cells analyzed per embryo per time point).

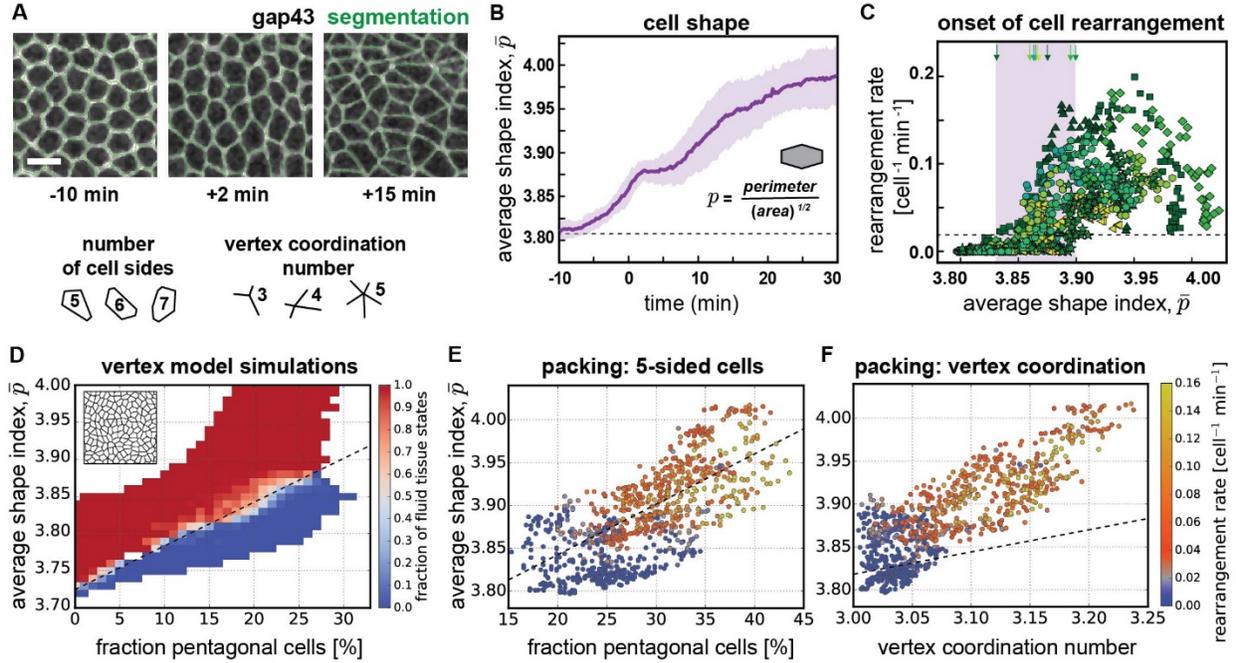

**Figure 2. Cell shape and packing disorder alone are not sufficient to predict the onset of cell rearrangements in the *Drosophila* germband.** (*A*) Confocal images from time lapse movies of epithelial cell patterns in the ventrolateral region of the germband tissue during *Drosophila* axis elongation. Cell outlines visualized using the fluorescently-tagged cell membrane marker, gap43:mCherry (53). Anterior left, ventral down. Images with overlaid polygon representations used to quantify cell shapes (green). Scale bar, 10 μm. See *SI Appendix* Fig. S1. (*B*) The average cell shape index $\bar{p}$ in the germband before and during convergent extension. The cell shape index, $p$, is calculated for each cell from the ratio of cell perimeter to square root of cell area, and the average value for cells in the tissue, $\bar{p}$, is calculated at each time point. The mean and standard deviation between embryos is plotted. Dashed line denotes the previously reported value for the solid-fluid transition in the isotropic vertex model, $\bar{p} = 3.81$. See also *SI Appendix* Fig. S2. (*C*) The instantaneous rate of cell rearrangements per cell versus the average cell shape index $\bar{p}$ from movies of individual embryos at time points before and during convergent extension in 8 wild-type embryos (different symbols correspond to different embryos). Small green arrows indicate the values of $\bar{p}$ at the onset of rapid cell rearrangement (>0.02 per cell per min, dashed line) in different embryos. Shaded region denotes values of $\bar{p}$ for which different embryos display distinct behaviors, either showing rapid cell rearrangement or not. Thus, a fixed value of $\bar{p}$ is not sufficient to determine the onset of rearrangement. (*D*) In vertex model simulations, the solid-fluid transition depends on exactly how cells are packed in the tissue (*SI Appendix*, Materials and Methods and Fig. S3). In model tissues, we find a linear dependence of the critical cell shape index on the fraction of pentagonal cells $f_5$, which is a metric for packing disorder. The dashed line represents a linear fit to this transition: $p_o^* = 3.725 + 0.59 f_5$ . (*E*) The relationship between $\bar{p}$ and $f_5$ for 8 wild-type embryos, with each point representing a time point in a single embryo. The dashed line is the prediction from vertex model results (same as in panel D). (*F*) The relationship between $\bar{p}$ and vertex coordination number for 8 wild-type embryos, with each point representing a time point in a single embryo. The dashed line is the prediction from Ref. (42). (*E-F*) Instantaneous cell rearrangement rate per cell in the tissue is represented by the color of each point, with blue indicating low rearrangement rates and red to yellow indicating high rearrangement rates.

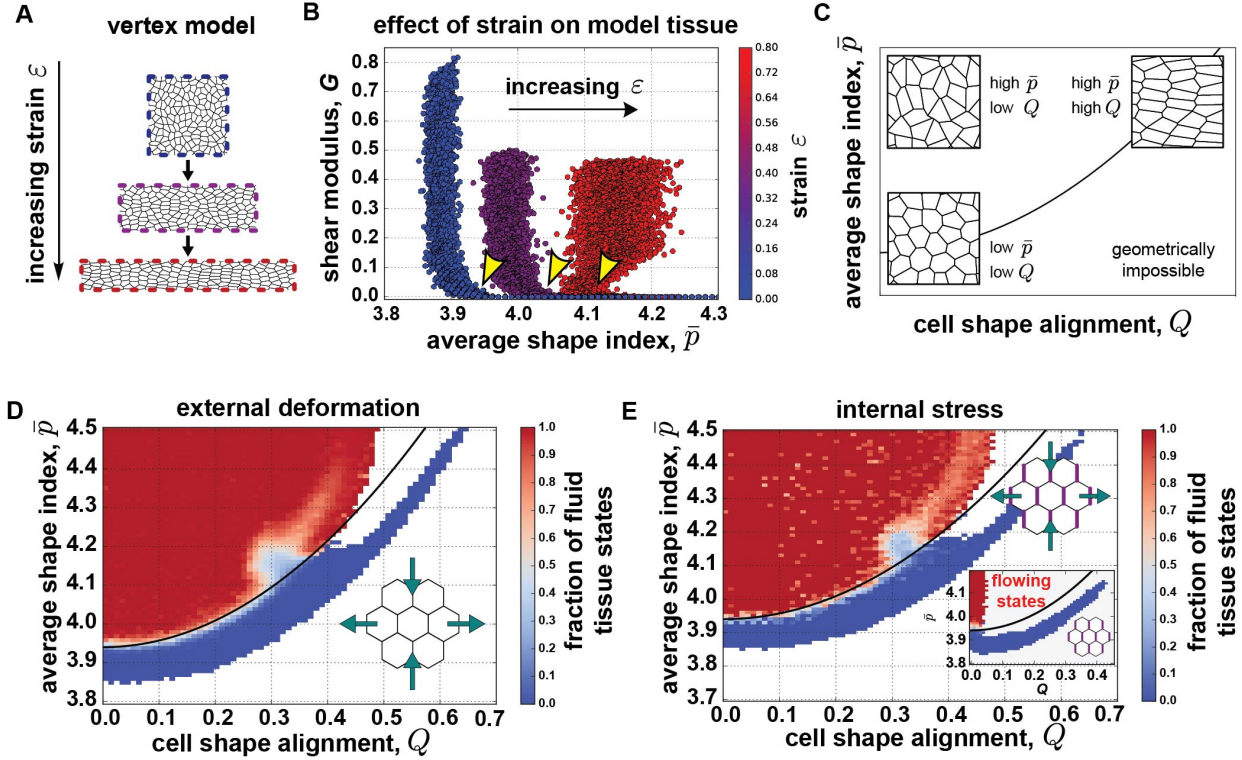

**Figure 3. The solid-to-fluid transition in a vertex model of anisotropic tissues.** (*A*) We study the effect of anisotropies on the solid-fluid transition in the vertex model by externally applying an anisotropic strain ε. An initially quadratic periodic box with dimensions $L_0 \times L_0$ is deformed into a box with dimensions $e^ε L_0 \times e^{-ε} L_0$. (*B*) Vertex model tissue rigidity as a function of the average cell shape index with different levels of externally applied strain ε (values for ε, increasing from blue to red: 0, 0.4, 0.8). For comparison, the strain in the wild-type germband between the times $t = 0$ min and $t = 20$ min is ε ≈ 0.6. For every force-balanced configuration, the shear modulus was analytically computed as described in the *SI Appendix*, SI Materials and Methods. For zero strain, we find a transition at an average cell shape index of $\bar{p}$ = 3.94 from solid behavior to fluid behavior. For increasing strain, the transition from solid to fluid behavior (i.e. the shear modulus becomes zero for a given strain) occurs at higher $\bar{p}$ (approximate positions marked by yellow arrows). Thus, a single critical cell shape index is not sufficient to determine the solid-fluid transition in an anisotropic tissue. (*C*) Cell shape and cell shape alignment can be used to characterize cell patterns in anisotropic tissues. Cell shape alignment $Q$ characterizes both cell shape anisotropy and cell shape alignment across the tissue. While a high cell shape index $\bar{p}$ correlates with anisotropic cell shapes, the cell shape alignment $Q$ is only high if these cells are also aligned. Conversely, low $\bar{p}$ implies low cell shape anisotropy and thus low $Q$. (*D-E*) Vertex model simulations for the case of an anisotropic tissue arising (*D*) due to externally induced deformation (cf. panels *A* and *B*), (*E*) due to internal active stresses generated by an anisotropic cell-cell interfacial tension combined with externally applied deformation, and (*E, inset*) due to internal active stresses without any externally applied force (*SI Appendix*). The fraction of tissue configurations that are fluid is plotted as a function of $\bar{p}$ and $Q$. For both internal and external sources of anisotropy, the critical shape index $\bar{p}$ marking the transition between solid states (blue) and fluid states (red) is predicted to depend quadratically on $Q$. White regions denote combinations

of $\bar{p}$ and $Q$ for which we did not find force-balanced states. In particular, in the case of finite tension anisotropy, we did not find any stable force-balanced fluid states, and the red fluid states in panel E all correspond to the limiting value of zero tension anisotropy. In *SI Appendix*, Materials and Methods we explain how the lack of fluid states for finite tension anisotropy can be explained analytically. Our findings quite generally suggest that stationary states of fluid tissues with an anisotropic cell-cell interfacial tension are difficult to stabilize even when preventing overall oriented tissue flow via the boundaries. In panel *D*, the solid line shows a fit of the transition to Eq. (1) with $p_o^* = 3.94$ and $b = 0.43$; panel *E* and *inset* show this same line. In panel *D*, a deviation from Eq. (1) is only seen around $\bar{p} \approx 4.15$ and $Q \approx 0.3$, where we observe an abundance of solid states, which is likely due to the occurrence of manyfold vertices in this regime (*SI Appendix* Fig. S4), which are known to rigidify vertex model tissue (42).

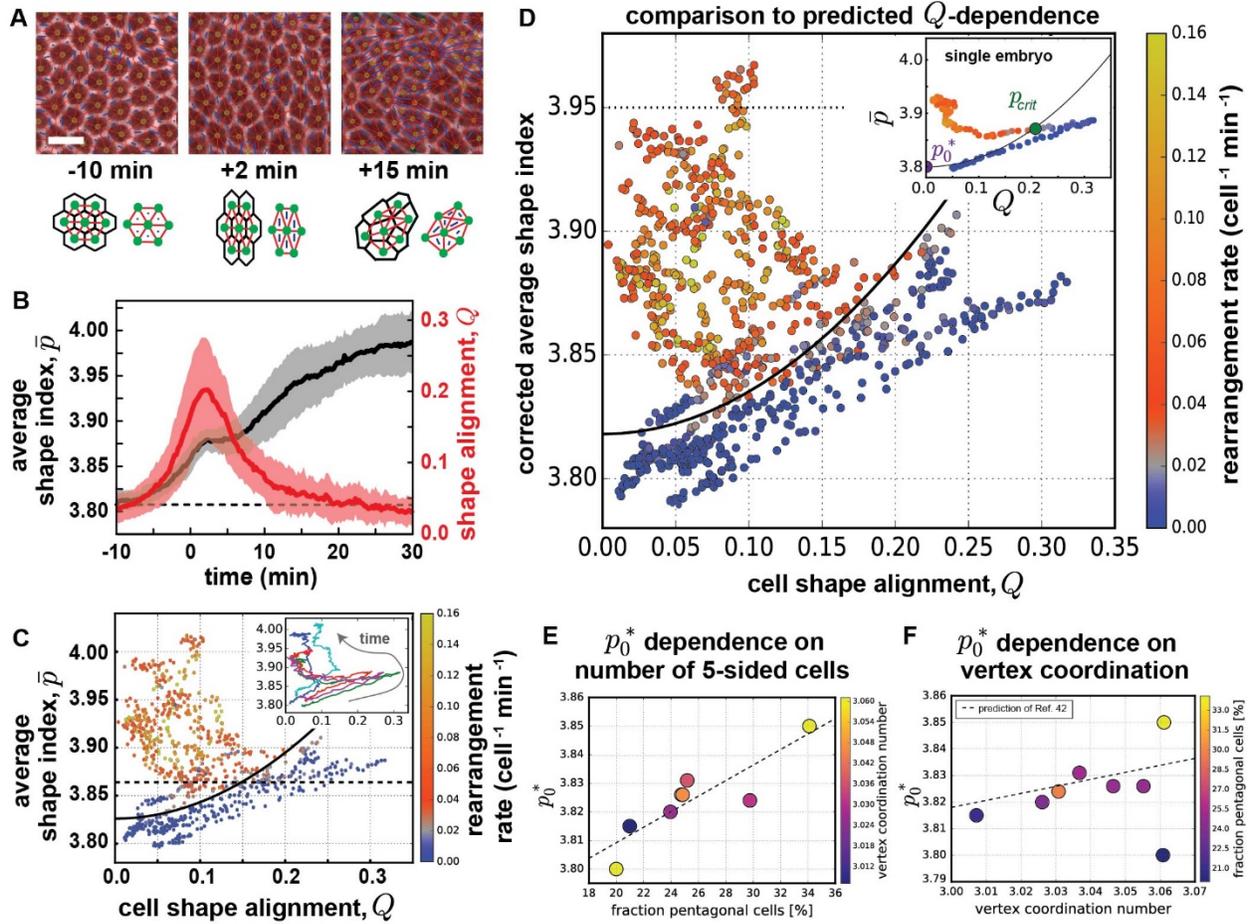

**Figure 4. Cell shape and cell shape alignment together predict the onset of cell rearrangements during *Drosophila* convergent extension.** (*A*) Confocal images from time lapse movies of epithelial cell patterns in the ventrolateral region of the germband during *Drosophila* axis elongation. Cell outlines were visualized with gap43:mCherry (53). Anterior left, ventral down. Scale bar, 10 μm. Images of cells with overlaid triangles that were used to quantify cell shape anisotropy. Cell centers (green dots) are connected with each other by a triangular network (red bonds). Cell shape stretches are represented by triangle stretches (blue bars), and the average cell elongation, $Q$, is measured (56). (*B*) The cell shape alignment index $Q$ (red) and average cell shape index $\bar{p}$ (black, same as Fig. 2*B*) for the germband tissue before and during axis elongation. $Q$ was calculated for each time point, and the mean and standard deviation between embryos is plotted (*N*=8 embryos with an average of 306 cells analyzed per embryo per time point). The onset of tissue elongation occurs at $t = 0$. The dashed line denotes the previously reported value for the solid-fluid transition in the isotropic vertex model, $\bar{p}$ =3.81 (39). (*C*) The relationship between $\bar{p}$ and $Q$ for 8 individual wild-type embryos, with each point representing $\bar{p}$ and $Q$ for a time point in a single embryo. Instantaneous cell rearrangement rate per cell in the tissue is represented by the color of each point, with blue indicating low rearrangement rates and red to yellow indicating high rearrangement rates. The black solid line indicates a fit to Eq. (1) with a rearrangement rate cutoff of 0.02 min⁻¹ per cell (*SI Appendix*, Materials and Methods), from which we extract $p_o^* = $ 3.83, where $b$ was fixed to the value obtained in vertex model simulations (cf. Fig. 3*D*). *Inset:* $\bar{p}$ and $Q$ for individual embryos over time. (*D*) The relationship between the corrected average cell

shape index $\bar{p}_{corr}$ and cell shape alignment $Q$ for 8 individual wild-type embryos, with each point representing a time point in a single embryo. The cell shape index is corrected by the vertex coordination number $z$ as $\bar{p}_{corr} = \bar{p} - (z-3)/B$, with $B = 3.85$ (42). Instantaneous cell rearrangement rate per cell in the tissue is represented by the color of each point. The solid line indicates the parameter-free prediction of Eq. (2). *Inset:* Single embryo fit to Eq. (1). *(E)* $p_o^*$ from single embryo fits to Eq. (1) correlate with the fraction of pentagonal cells $f_5$, a metric for cell packing disorder in the tissue, at the transition point. The dashed line represents a linear fit to the data. When using a rearrangement rate cutoff of 0.02 min⁻¹ per cell for the single embryo fits, we obtain for this linear fit: $p_o^* = 3.755 + 0.27f_5$ . *(F)* $p_o^*$ from single embryo fits to Eq. (1) correlate with the average vertex coordination number, another metric for packing disorder in the tissue, at the transition point. The dashed line represents the previous theoretical prediction for how manyfold vertices influence tissue behavior (42).

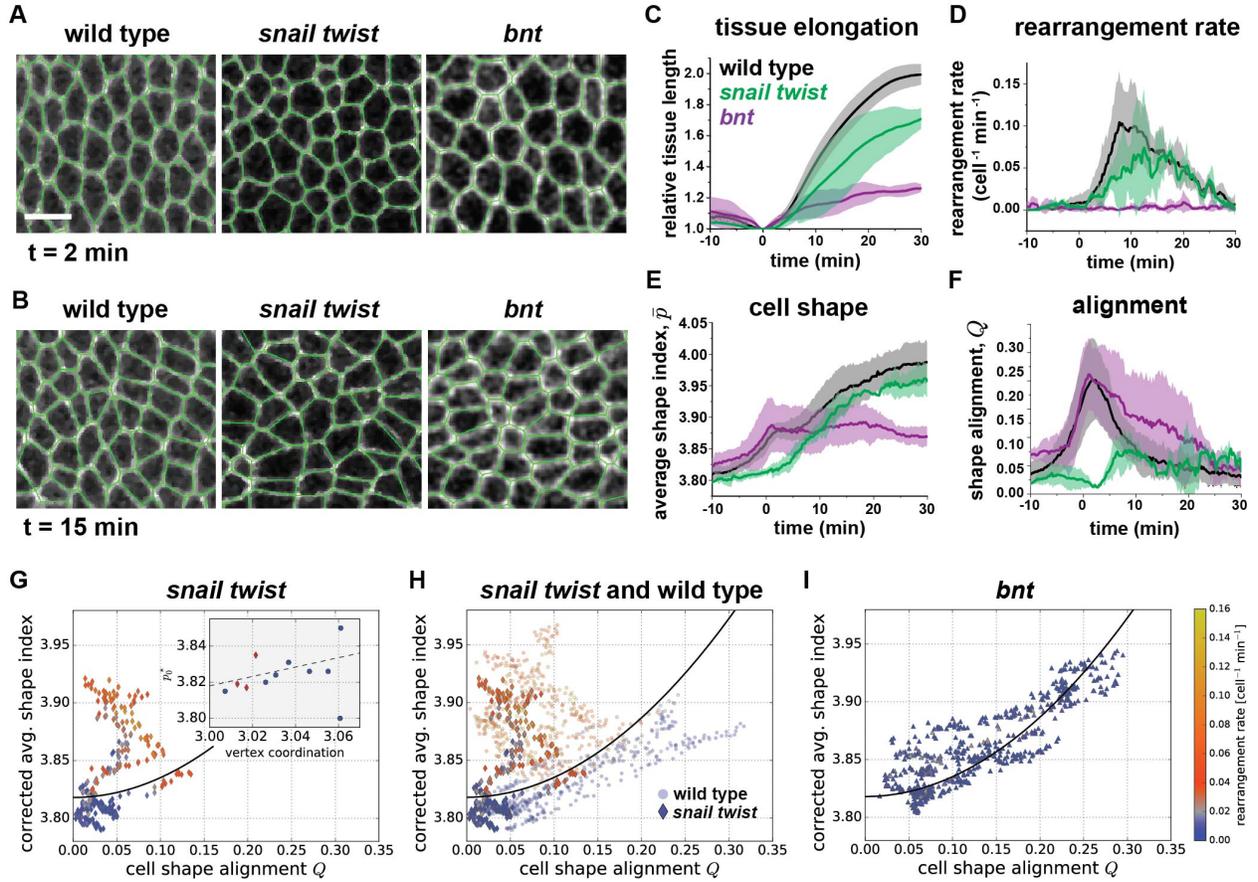

**Figure 5. Cell shape, cell shape alignment, and cell rearrangement rates in the germband of *snail twist* and *bnt* mutant embryos.** *snail twist* embryos lack ventral patterning genes required for presumptive mesoderm invagination. *bcd nos tsl* (*bnt*) embryos lack anterior-posterior patterning genes required for axis elongation and show severely disrupted myosin planar polarity compared to wild type (*SI Appendix* Fig. S10). (*A, B*) Confocal images from time lapse movies of cell patterns at $t = +2$ min and $t = +15$ min. Cell outlines visualized with fluorescently-tagged cell membrane markers: gap43:mCherry in wild type, Spider:GFP in *snail twist*, and Resille:GFP in *bnt*. Polygon representations of cell shapes are overlaid (green). Scale bar, 10 μm. (*C*) Tissue elongation is moderately reduced in *snail twist* and severely reduced in *bnt* compared to wild type. (*D*) Cell rearrangement rate is moderately decreased in *snail twist* and severely reduced in *bnt*. (*E*) In *snail twist*, the average cell shape index $\bar{p}$ is reduced compared to in wild type for -5 min < $t$ < 5 min. In *bnt*, $\bar{p}$ shows similar behavior to in wild-type for $t$ < 5 min, but does not show further increases with time for $t$ > 5 min. (*F*) In *snail twist*, the cell alignment index $Q$ is strongly reduced for -5 min < $t$ < 10 min compared to in wild type. In *bnt*, $Q$ shows similar behavior to in wild-type for $t$ < 5 min, but relaxes more slowly to low levels. (*C-F*) The mean and standard deviation between embryos is plotted (3 *snail twist* and 5 *bnt* embryos with an average of 190 cells per embryo per time point). (*G-I*) Relationship between the corrected cell shape index $\bar{p}_{corr}$ and $Q$ for 3 *snail twist* (*G,H*), 8 wild-type (*H*), and 5 *bnt* (*I*) embryos, with each point representing a time point in a single embryo. Instantaneous rearrangement rate is represented by the color of each point. Solid lines represent the prediction of Eq. (2). (*H*) Tissue behavior in *snail twist* and wild-type embryos, all of which exhibit rapid cell rearrangement during convergent extension, is well described by the prediction of Eq. (2), which does not require any fitting parameters.

## Supporting Information

## SI Materials and Methods

**Fly Stocks and Genetics.** Embryos were generated at 23°C and analyzed at room temperature. Wild-type control embryos were *yw* with one maternal copy of a sqh-gap43:mCherry transgene to label cell membranes (1). *snail twist* embryos were zygotic *snail*[IIG05] *twist*[DfS60] mutants and expressed Spider:GFP to visualize cell outlines (2, 3). The *bcd nos tsl* (*bnt*) maternal mutant embryos were the progeny of *bcd*[E1] *nos*[L7] *tsl*[146] homozygous females and expressed Resille:GFP to visualize cell outlines.

**Time-Lapse Imaging.** Embryos aged 2-4 hours were dechorionated for 2 min in 50% (vol/vol) bleach, washed in distilled water, and mounted in halocarbon oil 27 and 700, 1:1 (Sigma) between a coverslip and an oxygen-permeable membrane (YSI). Embryos were oriented with the cephalic furrow and ventral furrow just visible at the edges of the field of view (*SI Appendix*, Fig. S1). A 212 μm x 159 μm ventrolateral region of the embryo was imaged on a Zeiss LSM880 laser scanning confocal microscope with a 40X/1.2 NA water-immersion objective. Z-stacks were acquired at 1-μm steps and 15-s time intervals. Maximum intensity z-projections of 3 μm in the apical junctional plane were analyzed.

**Tissue Elongation Measurement.** Tissue elongation was measured by particle image velocimetry (PIV) using PIVlab in MATLAB (4). Each image was divided into 2-pass Fast-Fourier-Transform windows ($120 \times 120$ pixels) with 50% overlaps. A displacement vector field for each window and each time point was determined by cross-correlating each window in the current time point and the image in the next time point. Tissue length change was measured by quantifying the cumulative sum of the anterior-directed displacement at the anterior end of the germband and the posterior-directed displacement at the posterior end of the germband. The onset of tissue elongation ($t = 0$) was the time point when the derivative of the tissue elongation curve intersects zero.

**Automated Image Segmentation and Cell Rearrangement Analysis.** Time-lapse movies were projected and despeckled using ImageJ. Processed movies were segmented and computationally analyzed using the MATLAB based software SEGGA, and errors were corrected manually with the interactive user interface (3). Cells were tracked and analyzed between $t = -10$ min and $t = 30$ min for each movie. To be included in cell rearrangement analysis, cells must be in the region of interest for at least 5 minutes after $t = 0$. The cell rearrangement rate shown is a uniformly weighted average over 1.5 minutes.

**Cell Shape Index and Cell Shape Alignment Analysis.** Based on the cell segmentation data, we computed the average cell shape index $\bar{p}$ by quantifying for each segmented cell both the perimeter $P$ and area $A$ of the polygon defined by the cell vertices (i.e. the points where at least 3 cells meet). The average cell shape index $\bar{p}$ at a time point is the average of $P/\sqrt{A}$ over all segmented cells.

Cell shape alignment $Q$ was quantified using the triangle method following Ref. (5). A triangular tiling was created based on the barycenters of the cellular polygons, where each vertex of the cellular network gives rise to a triangle whose corners are defined by the barycenters of the three abutting cells. In the case of a manyfold vertex, i.e. a vertex abutting $M > 3$ cells, $M$ triangles are created, where each triangle has one corner defined as the average position of the barycenters of all $M$ abutting cells and the other two corners are the barycenters of two adjacent cells. For each triangle, we computed a symmetric, traceless tensor $\boldsymbol{q}$ quantifying triangle elongation. To compute the tensor $\boldsymbol{q}$ for a given triangle $m$, we first define a shape tensor $\boldsymbol{s}$ that corresponds to the affine deformation transforming an equilateral reference triangle into the observed triangle $m$. With the corners of the triangle being at positions $\boldsymbol{r}^A$, $\boldsymbol{r}^B$, $\boldsymbol{r}^C$ in counter-clockwise order, the shape tensor $\boldsymbol{s}$ can be computed as follows:

$$\boldsymbol{s} = \begin{pmatrix} r_x^B - r_x^A & r_x^C - r_x^A \\ r_y^B - r_y^A & r_y^C - r_y^A \end{pmatrix} \cdot \begin{pmatrix} 1 & 1/2 \\ 0 & \sqrt{3}/2 \end{pmatrix}^{-1}. \tag{S1}$$

From the triangle shape tensor, the triangle elongation tensor $\boldsymbol{q}$ is extracted, which characterizes the anisotropic component of the deformation characterized by $\boldsymbol{s}$. It is computed by first splitting $\boldsymbol{s}$ into trace part $\boldsymbol{t}$, symmetric, traceless part $\tilde{\boldsymbol{s}}$, and antisymmetric part $\boldsymbol{s}^a$:

$$\boldsymbol{s} = \boldsymbol{t} + \tilde{\boldsymbol{s}} + \boldsymbol{s}^a. \tag{S2}$$

Then first a triangle rotation angle $\theta$ is extracted such that:

$$\begin{pmatrix} \cos\theta \\ \sin\theta \end{pmatrix} = a \begin{pmatrix} t_{xx} \\ s_{yx}^a \end{pmatrix} \tag{S3}$$

with some prefactor $a$. In practice, $\theta$ can be extracted using the "arctan2" function that exists in many programming languages as $\theta = \arctan2(s_{yx}^a, t_{xx})$. Finally, the triangle elongation tensor $\boldsymbol{q}$ is computed as:

$$\boldsymbol{q} = \frac{1}{|\tilde{\boldsymbol{s}}|} \operatorname{arcsinh}\left(\frac{|\tilde{\boldsymbol{s}}|}{(\det \boldsymbol{s})^{1/2}}\right) \tilde{\boldsymbol{s}} \cdot \boldsymbol{R}(-\theta), \tag{S4}$$

where $|\tilde{\boldsymbol{s}}| = \left[s_{xx}^2 + s_{xy}^2\right]^{1/2}$ is the magnitude of the symmetric, traceless tensor $\tilde{\boldsymbol{s}}$, $\det \boldsymbol{s}$ is the determinant of the shape tensor $\boldsymbol{s}$, and $\boldsymbol{R}(-\theta)$ is a clockwise rotation by angle $\theta$:

$$\boldsymbol{R}(-\theta) = \begin{pmatrix} \cos\theta & \sin\theta \\ -\sin\theta & \cos\theta \end{pmatrix}. \tag{S5}$$

The cell shape alignment tensor

$$\boldsymbol{Q} = \begin{pmatrix} Q_{xx} & Q_{xy} \\ Q_{xy} & -Q_{xx} \end{pmatrix} \tag{S6}$$

is then the average of the symmetric, traceless elongation tensors $\boldsymbol{q}$ of all triangles:

$$\boldsymbol{Q} = \langle \boldsymbol{q} \rangle. \tag{S7}$$

The average is an area-weighted average $\langle \boldsymbol{q} \rangle := (\sum_m a_m \boldsymbol{q}_m)/(\sum_m a_m)$, where the sums are over all triangles $m$ with area $a_m$ and elongation tensor $\boldsymbol{q}_m$. The cell shape alignment parameter $Q$ in the main text is the magnitude of this tensor defined by $Q = \left[Q_{xx}^2 + Q_{xy}^2\right]^{1/2}$.

Our cell shape alignment parameter $Q$ combines information about both cell shape anisotropy and cell shape alignment. It can be split accordingly into a product:

$$Q = |\langle \boldsymbol{q} \rangle| = Q_s Q_a. \tag{S8}$$

The first factor $Q_s = \langle |\boldsymbol{q}| \rangle$ is the average magnitude of triangle anisotropy, which is a proxy for cell shape anisotropy, and the second factor $Q_a = \left| \langle \frac{\boldsymbol{q}}{|\boldsymbol{q}|} \frac{|\boldsymbol{q}|}{\langle |\boldsymbol{q}| \rangle} \rangle \right|$ is the norm of the average triangle elongation axis $\boldsymbol{q}/|\boldsymbol{q}|$ weighted by the norm of $\boldsymbol{q}$. This second factor thus corresponds to an alignment separate from cell shape, which similarly to a nematic order parameter varies between zero (random shape orientation) and one (perfectly aligned shapes).

**Vertex Model.** Our vertex model describes an epithelial tissue as a planar tiling of $N$ cellular polygons, where the degrees of freedom are the vertex positions $r_{k\alpha}$ (6). We use Latin indices starting with $k$ to refer to vertices and Greek indices starting with $\alpha$ to refer to spatial dimensions. Forces are defined such that cell perimeters and areas act as effective springs with a preferred perimeter $p_0$ and a preferred area of one. This is implemented via the following effective energy functional, which in dimensionless form is (7):

$$E = \sum_{i=1}^{N}[(p_i - p_0)^2 + k_A(a_i - 1)^2] \tag{S9}$$

Here, the sum is over all cells $i$, with perimeter $p_i$ and area $a_i$. The parameter $k_A$ is a dimensionless number comparing area and perimeter rigidity. We use periodic boundary conditions with box size $L_x \times L_y$ such that the average cell number density is one: $L_x L_y = N$. The boundary conditions can accommodate a skew (as in Lees-Edwards boundary conditions) with a corresponding simple shear $\gamma$. Hence, the system energy is a function of all vertex positions and the periodic box parameters: $E = E(\{r_{k\alpha}\}, L_x, L_y, \gamma)$. We focus on stable, force-balanced states of the system, which corresponds to local minima of $E$. To numerically find such states, we use the BFGS2 multidimensional minimization routine of the Gnu scientific library (GPL) with a cutoff on the average residual force of $10^{-6}$. We allow for manyfold vertices – i.e. vertices are allowed to be in contact with more than three cells at once. During the minimization, the vertices belonging to an edge are fused to a single vertex whenever the edge length is below a cutoff of $10^{-3}$, and a vertex with at least four edges attached to it splits into several vertices whenever this is energetically favorable. While it is known that the existence of manyfold vertices can change the transition point in vertex models (8), we checked that the energy-minimized states we obtained rarely contained any manyfold vertices. In all simulations, we have $N = 512$ cells and $k_A = 1$.

For a given local energy minimum, we compute the simple shear modulus $G$ as described in (9):

$$G = \frac{1}{N} \left( \frac{\partial^2 E}{\partial \gamma^2} - \sum_m \frac{1}{\omega_m^2} \left[ \sum_{k,\alpha} \frac{\partial^2 E}{\partial \gamma \partial r_{k\alpha}} u_{k\alpha}^m \right]^2 \right) \tag{S10}$$

In the second term, the outer sum is over all positive eigenvalues $\omega_m^2$ and the corresponding eigenvectors $u_{k\alpha}^m$ of the Hessian matrix $(\partial^2 E / \partial r_{k\alpha} \partial r_{l\beta})$. In practice, we include all eigenvalues smaller than $10^{-14}$ in the sum. The inner sum in the second term is over all vertices and both spatial dimensions.

***Anisotropic vertex model.*** In all our simulations, we initialize the system with the Voronoi tessellation of a uniformly random point pattern on a squared domain ($L_x = L_y = L_0$). For the first set of simulations of anisotropic tissue (Fig. 3$D$), we apply an external pure shear strain $\varepsilon$ by setting $L_x = e^\varepsilon L_0$ and $L_y = e^{-\varepsilon} L_0$. We start with $\varepsilon = 0$ and increase in steps of $0.02$ up to a value of $\varepsilon = 2$, minimizing the energy after each step. For these minimizations, we vary all vertex positions, keep the box dimensions fixed, but also allow the simple shear variable $\gamma$ to vary (shear-stabilized minimization). We follow this protocol for different values of $p_0$, which we varied between 3.5 and 4.5 in steps of 0.01. For each value of $p_0$ we carry out 100 separate simulation runs.

For a second set of simulations of an anisotropic tissue (Fig. 3*E*, *inset*), we model the anisotropic myosin distribution in the germband by introducing an additional anisotropic line tension with amplitude $\lambda_0$ into the effective energy functional:

$$E = \sum_{i=1}^{N}[(p_i - p_0)^2 + k_A(a_i - 1)^2] + \sum_{\langle k,l \rangle} \lambda_{\langle k,l \rangle} \ell_{\langle k,l \rangle} \qquad (S11)$$

While the first sum is the same as in Eq. (S9), we have added a second sum, which is over all edges in the system, connecting two vertices $k, l$. Here, $\ell_{\langle k,l \rangle}$ denotes the length the edge $\langle k, l \rangle$, and $\lambda_{\langle k,l \rangle}$ is a line tension associated with this edge. Before each minimization, we define each of these line tensions based on the respective edge angle $\theta_{\langle k,l \rangle}$ as follows:

$$\lambda_{\langle k,l \rangle} = \lambda_0 \cos(2[\theta_{\langle k,l \rangle} - \phi]) \qquad (S12)$$

Thus, the line tension will be increased by $\lambda_0$ for edges parallel to lines with angle $\phi$ and decreased by $\lambda_0$ for edges perpendicular to that.

During each minimization run, we vary all vertex positions and the pure shear strain $\varepsilon = 1/2 \log(L_x/L_y)$, but keep the system area $L_x L_y$ and the simple shear strain $\gamma$ fixed. While we set $\lambda_{\langle k,l \rangle}$ before a minimization run and keep it constant during the minimization, the angle $\theta_{\langle k,l \rangle}$ usually changes during the minimization as the vertex positions are varied. As a consequence, the state obtained after the minimization will not correspond to an energy minimum anymore once we update the line tensions $\lambda_{\langle k,l \rangle}$ with the new angles $\theta_{\langle k,l \rangle}$. Thus, to identify a force-balanced state where the line tensions are consistent with the directions of the cell edges, we iterate over several minimizations, where after each minimization we update $\lambda_{\langle k,l \rangle}$ based on the latest angles $\theta_{\langle k,l \rangle}$. We stop these iterations once the states do not significantly change anymore, or more precisely, when the average residual stress per degree of freedom *before* a minimization, but with the *new* $\lambda_{\langle k,l \rangle}$, is smaller than $2 \times 10^{-6}$. We intentionally *do not* include the explicit dependency of $\lambda_{\langle k,l \rangle}$ on $\theta_{\langle k,l \rangle}$ and thus the vertex positions in our energy minimizations, because this would create additional torques in our model, while here we merely want to study the effect of an anisotropic distribution of line tensions as provided for instance by an anisotropic myosin distribution.

We set the direction of line tension anisotropy parallel to the $y$ axis, i.e. $\phi = \pi/2$. For Fig. 3*E*, we varied the magnitude of line tension anisotropy $\lambda_0$ between zero and one in steps of 0.1, and for Fig. 3*E* inset, we varied it in steps of 0.01. Again, the preferred perimeter $p_0$ is varied between 3.5 and 4.5 in steps of 0.01, where for each value of $p_0$ we run 100 separate simulations.

We found many states where the system flowed during a minimization until $\varepsilon$ was so large that the system was only one cell thick in the $y$ direction. In particular, this was the case in what was otherwise expected to be the floppy regime. This will probably not only occur in the floppy regime, but also in the solid regime whenever the anisotropic stress created by the line tension anisotropy is large enough to overcome the yield stress, which perhaps explains why there is a gap between mechanically stable solid states and the black line in Fig. 3*E* inset. Because we did not obtain any force-balanced state of bulk vertex model tissue in this regime, we have no way to determine from our simulations neither the shear modulus, nor the morphological quantities $\bar{p}$ and $Q$, and thus this regime does not appear in Fig. 3$E$ inset. To access this regime, one needs to include dynamics into the model, e.g. including a viscosity or a substrate friction.

In order to prevent the system from flowing indefinitely, we also ran a third set of simulations (Fig. 3$E$), where we combined anisotropic line tensions with a fixed system size. In other words, we ran simulations like the second set where we now fixed also the pure shear strain $\varepsilon$, which we successively increased and each time looked for a force-balanced state as described for the second simulation set. While our findings are consistent with our results from the other two simulation sets, we again do not obtain any fluid states for any nonzero value of the line tension anisotropy $\lambda_0$, because the iterative procedure involving updating line tensions and minimizing the energy described above does not converge in this regime. To some extent this can even be understood analytically. We start from a fix point $(\mathbf{r}^*, \boldsymbol{\theta}^*) = (\{\vec{r}_k^*\}, \{\theta_{(k,l)}^*\})$ of the iterative procedure described above, where $\partial E/\partial \mathbf{r}(\mathbf{r}^*, \boldsymbol{\theta}^*) = 0$ and $\boldsymbol{\theta}^* = \boldsymbol{\theta}(\mathbf{r}^*)$. A small deviation $\boldsymbol{\delta r}_1$ from this state in the vertex positions leads to a change in the bond angles of $\boldsymbol{\delta \theta}_1 = \mathrm{d}\boldsymbol{\theta}(\mathbf{r}^*)/\mathrm{d}\mathbf{r} \cdot \boldsymbol{\delta r}_1$. This leads in turn to an energy minimized state at vertex positions $\mathbf{r}_2 = \mathbf{r}^* + \boldsymbol{\delta r}_2$ with

$$0 = \frac{\partial E(\mathbf{r}^* + \boldsymbol{\delta r}_2, \boldsymbol{\theta}^* + \boldsymbol{\delta \theta}_1)}{\partial \mathbf{r}} = \frac{\partial^2 E(\mathbf{r}^*, \boldsymbol{\theta}^*)}{\partial \mathbf{r}^2} \cdot \boldsymbol{\delta r}_2 + \frac{\partial^2 E(\mathbf{r}^*, \boldsymbol{\theta}^*)}{\partial \mathbf{r} \partial \boldsymbol{\theta}} \cdot \frac{d\boldsymbol{\theta}(\mathbf{r}^*)}{d\mathbf{r}} \cdot \boldsymbol{\delta r}_1. \quad (S13)$$

Here, $\partial^2 E(\mathbf{r}^*, \boldsymbol{\theta}^*)/\partial \mathbf{r}^2$ is the Hessian of the system and $\partial^2 E(\mathbf{r}^*, \boldsymbol{\theta}^*)/\partial \mathbf{r} \partial \boldsymbol{\theta}$ scales linearly with the line tension anisotropy $\lambda_0$. According to Eq. (S13), if eigenvalues of the Hessian are sufficiently small that have a nonzero overlap with the second term in Eq. (S13), then any initially small deviation $\boldsymbol{\delta r}_1$ from the fix point $(\mathbf{r}^*, \boldsymbol{\theta}^*)$ will grow during the iterative procedure. In particular, this would explain why we did not observe any stable fluid states for $\lambda_0 > 0$, where the Hessian is expected to contain non-trivial zero modes. We note that this is likely more than just a technical phenomenon related to our quasi-static simulations. This result could instead indicate that generally fluid tissues with anisotropic line tension may not be able to easily attain a stationary state even when boundary conditions prevent overall anisotropic tissue flow.

To obtain Fig. 3$D$,$E$ & inset we binned all of our energy-minimized configurations with respect to $\bar{p}$ and $Q$ (which were computed as described above), and then computed the fraction of floppy configurations within each bin. A configuration was defined floppy when its shear modulus $G$ was below a cutoff value of $10^{-5}$.

***Packing dependence of transition point.*** To study the packing-dependence of the transition point, we annealed the isotropic vertex model tissue at different temperatures prior to quenching the system to zero temperature, as this is a standard method for altering packing disorder in other materials such as structural glasses. To simulate the vertex model at a given temperature, we followed an Euler integration scheme updating all vertex positions $r_{k\alpha}$ as follows in each time step $\Delta t$:

$$r_{k\alpha} \rightarrow r_{k\alpha} + \mu F_{k\alpha} \Delta t + \eta_{k\alpha} \quad (S14)$$

Here, we have non-dimensionalized time such that the dimensionless motility $\mu$ is one, $F_{k\alpha} = -\partial E/\partial r_{k\alpha}$ is the force on vertex $k$ with the energy given by Eq. (S9), and $\eta_{k\alpha}$ is a normal distributed random force with zero average and variance $\langle \eta_{k\alpha} \eta_{l\beta} \rangle = 2\mu T \Delta t \delta_{kl} \delta_{\alpha\beta}$. To simulate these dynamics, we use the publicly available cellGPU code (10), with a time step of $\Delta t = 0.01$. For these simulations, vertices are always 3-fold coordinated and an edge undergoes a full T1 transition whenever its length is below a cutoff of 0.04.

We run 100 simulations for each set of parameters $(T, p_0)$, where $T$ varies logarithmically between $5 \times 10^{-6}$ and $1.5 \times 10^{-1}$, and $p_0$ varies between 3.7 and 3.9 in steps of 0.01. All simulations are thermalized at their target temperature for a time of $10^4$ before recording the data. We then perform simulations for $10^6$ at the target temperature before the temperature is quenched to $T = 0$. We calculate the decay of the self-overlap function for the slowest of the most solid states (low $p_0$ and $T$ sets) and confirm that the vertices are displaced less than a characteristic distance of $1/e$ at time $10^6$, suggesting that the states are relaxed. To quench the temperature to zero, we run Eq. (S14) for an additional time of $10^6$. Afterwards, we use the BFGS2 algorithm of the GSL to further minimize until the average residual force per degree of freedom is below $10^{-6}$. The data are shown in *SI Appendix*, Fig. S3, demonstrating that the transition point $p_0{}^*$ does depend systematically on the annealing temperature and therefore on the packing disorder.

The transition point we find occasionally decreases below the value of 3.81 (*SI Appendix* Fig. S3), which is the minimal transition point we would expect for disordered packings (11). To test whether this could be due to partial crystallization, we also quantified a hexatic order parameter:

$$\Phi_6 = \frac{1}{N_e} \sum_{\langle k,l \rangle} e^{6i\theta_{\langle k,l \rangle}} \tag{S15}$$

Here, the sum is over all $N_e$ edges in the system, where $\theta_{\langle k,l \rangle}$ is the angle of the edge between vertices $k$ and $l$. We find that the decreased transition point is indeed correlated with hexatic order $|\Phi_6|^2$ (*SI Appendix*, Fig. S3).

**Theoretical Expectation for the Shift of the Transition Point.** In a recent publication (7), some of us showed that the transition point $\bar{p}_{\text{crit}}$ in the vertex model is expected to shift away from the isotropic transition point $p_o^*$ as the material is anisotropically deformed with strain $\varepsilon = 1/2 \log(L_x/L_y)$ as:

$$\bar{p}_{\text{crit}} = p_o^* + 4b\varepsilon^2 \tag{S16}$$

Here, $b$ is a constant prefactor whose precise value depends on the packing disorder, but whose typical value was previously found to be 0.6±0.2 (average ± standard deviation). We use here the pure shear strain variable $\varepsilon = 1/2 \log(L_x/L_y)$, which is related to the strain variable $\gamma$ used in Ref. (7) as $\varepsilon = \gamma/2$, and so we get an additional factor of 4 in front of $b$ in Eq. (S16). However, Eq. (S16) has so far only been discussed without cell rearrangements, which do occur in our simulations.

To apply these ideas here, we start from an anisotropic configuration that results from some externally applied area-preserving anisotropic strain $\varepsilon_{xx}$ that for simplicity we define here to extend the tissue along the $x$ axis. It is possible that cell rearrangements occur during this deformation process, which are not taken into account in Eq. (S16). Thus, we need to disentangle overall strain from the cell rearrangements that it may cause. In Ref. (5), some of us have shown before that in the limit of homogeneous deformation without global rotations, the overall strain of a dense 2D cellular network can be decomposed into:

$$\varepsilon_{xx} = Q_{xx}^{\text{final}} - Q_{xx}^{\text{initial}} + \sum_i \Delta X_{xx}^i \tag{S17}$$

Here, $Q_{xx}^{\text{initial}}$ and $Q_{xx}^{\text{final}}$ are measures for nematic cell shape alignment before and after the deformation process, defined as described in section "Cell Shape Index and Cell Shape Alignment Analysis" (Eq. (S1) and following), projected on the $x$ axis. The sum is over all T1 transitions $i$ that occur during the deformation process, where each T1 transition contributes an amount $\Delta X_{xx}^i$ to the overall strain. Ref. (5) more generally derives a relation about symmetric, traceless tensors, which we projected onto the $x$ axis for simplicity here.

To apply Eq. (S16) to any anisotropic configuration with nematic cell shape alignment $Q_{xx}$, we ask for the amount of strain needed to deform this anisotropic configuration into an isotropic one *without* any cell rearrangements. Defining an isotropic configuration as one where the nematic cell shape alignment is zero $Q_{xx}^{\text{final}} = 0$, we obtain from Eq. (S17) that the strain needed to deform our starting configuration into an isotropic one without T1 transitions is $\varepsilon_{xx} = -Q_{xx}$. Conversely a strain of $\bar{\varepsilon}_{xx} = Q_{xx}$ is needed to get from that isotropic state back to the initial anisotropic state. With $p_o^*$ being the transition point of the isotropic state, we thus obtain from Eq. (S16) that the transition point of our anisotropic tissue state is:

$$\bar{p}_{\text{crit}} = p_o^* + 4b\bar{\varepsilon}_{xx}^2 = p_o^* + 4bQ_{xx}^2 \tag{S18}$$

Here we have assumed that all deformations are along the $x$ axis, but the same line of argument applies to any arbitrary axis, such that we finally obtain

$$\bar{p}_{\text{crit}} = p_o^* + 4bQ^2 \tag{S19}$$

Here, $Q$ being the magnitude of the nematic cell shape alignment defined in section "Cell Shape Index and Cell Shape Alignment Analysis" above.

An alternative way to obtain Eq. (S19) is to Taylor expand $\bar{p}_{\text{crit}}$ in terms of the cell shape alignment tensor $\boldsymbol{Q}$, where the lowest-order term besides the constant allowed by symmetry is a term $\sim Q^2$. However, with the approach above, we can also connect the value of the prefactor $b$ to previous results.

Finally, the predictions in Ref. (7) strictly speaking refer to the non-dimensionalized average perimeter, i.e. the average of $p_i$ over all cells, whereas here by $\bar{p}$ we refer to the average shape index, i.e. the average of $p_i/\sqrt{a_i}$ over all cells. We verified that this difference does not play a role in our vertex model simulations.

***Fit to simulation data.*** To fit Eq. (S19) to the simulation data where we apply the external anisotropic deformation (Fig. 3*D*), we compute the average transition point for each $Q$ by interpreting the $\bar{p}$-dependent fraction of floppy configurations for fixed $Q$ as a cumulative probability distribution function and extracting the average from it. For varying $Q$, the resulting plot together with a fit to Eq. (S19) is shown in *SI Appendix*, Fig. S4*A*.

We excluded a few data points from the fit, which were affected by the excess of rigid states observed around $\bar{p} \approx 4.15$ and $Q \approx 0.3$. This excess of rigid states very likely comes from the occurrence of higher coordinates vertices (*SI Appendix* Fig. S4*B*), which are known to increase the vertex model transition point (8).

***Fits to experimental data and quality of fit.*** To compare our theoretical predictions to experimental data, we define a quality of fit measure $n_{\text{tot}}$, which we define as the number of experimental data points that are wrongly categorized as either solid or fluid by our theory. Experimentally, a data point is declared fluid if the instantaneous cell rearrangement rate averaged over a 1.5 minute time interval surpasses a cutoff value, which we set to 0.02 rearrangements per cell and minute for the plots in the main text Figs. 2*C*, 4*C-F*, 5*G-I*. A few of our theoretical predictions include fit parameters. To determine their value from experimental data, we minimize the quality of fit measure $n_{\text{tot}}$ varying those fit parameters.

In *SI Appendix* Fig. S8, we compare several theoretical predictions by plotting the quality of fit over the rearrangement rate cutoff. To obtain a reliable measurement for the cell rearrangement rate, we fixed the lower limit for the rearrangement rate cutoff requiring that the standard error $2\sigma_r$ of the cell rearrangement rate that we measure is at most as big as its average $r$. Assuming a Poissonian distribution of the number $n_{T1}$ of cell rearrangements during the measurement interval of $\Delta t = 1.5$ min, we thus find:

$$1 \geq \frac{2\sigma_r}{r} = \frac{2\sigma_{n_{T1}}}{n_{T1}} = \frac{2}{\sqrt{n_{T1}}} = \frac{2}{\sqrt{r\Delta t N_c}}. \tag{S20}$$

Hence, to get a reliable measurement for the cell rearrangement rate, we choose a minimal cutoff value of $r_{\text{min}} = \frac{4}{\Delta t N_c} = 0.014$ min$^{-1}$ per cell, where $N_c$ is the total number of cells, which is on average $N_c = 190$.

**Statistical Analysis.** Unless otherwise noted, error bars are the standard deviation.

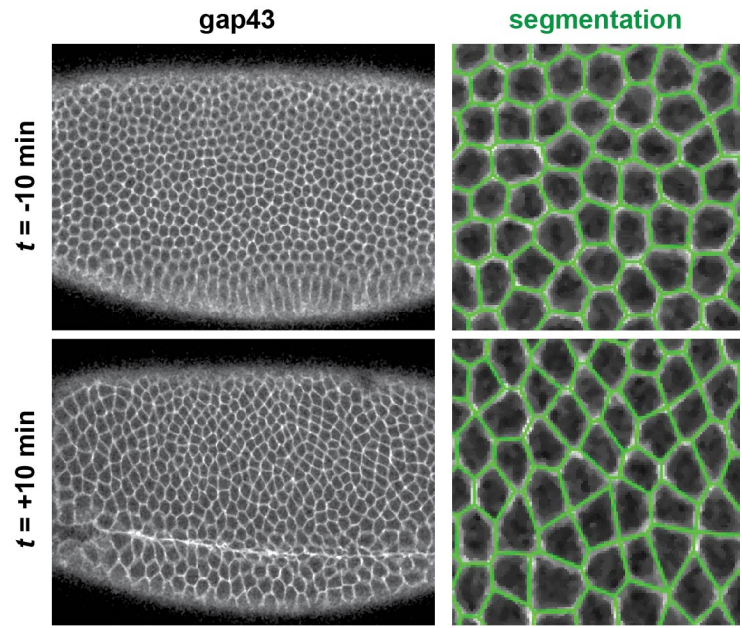

**Figure S1. Imaging and analysis of the germband epithelial tissue.** (*Left*) Confocal images from time lapse movies of epithelial cell patterns in the ventrolateral region of the germband tissue during *Drosophila* body axis elongation. Cell outlines were visualized using the fluorescently-tagged cell membrane marker, gap43:mCherry. Anterior left, ventral down. Images, 212 µm x 159 µm. (*Right*) Zoomed-in regions from images at left with overlaid polygon representations used to quantify cell shapes (green). Images, 40 µm x 40 µm.

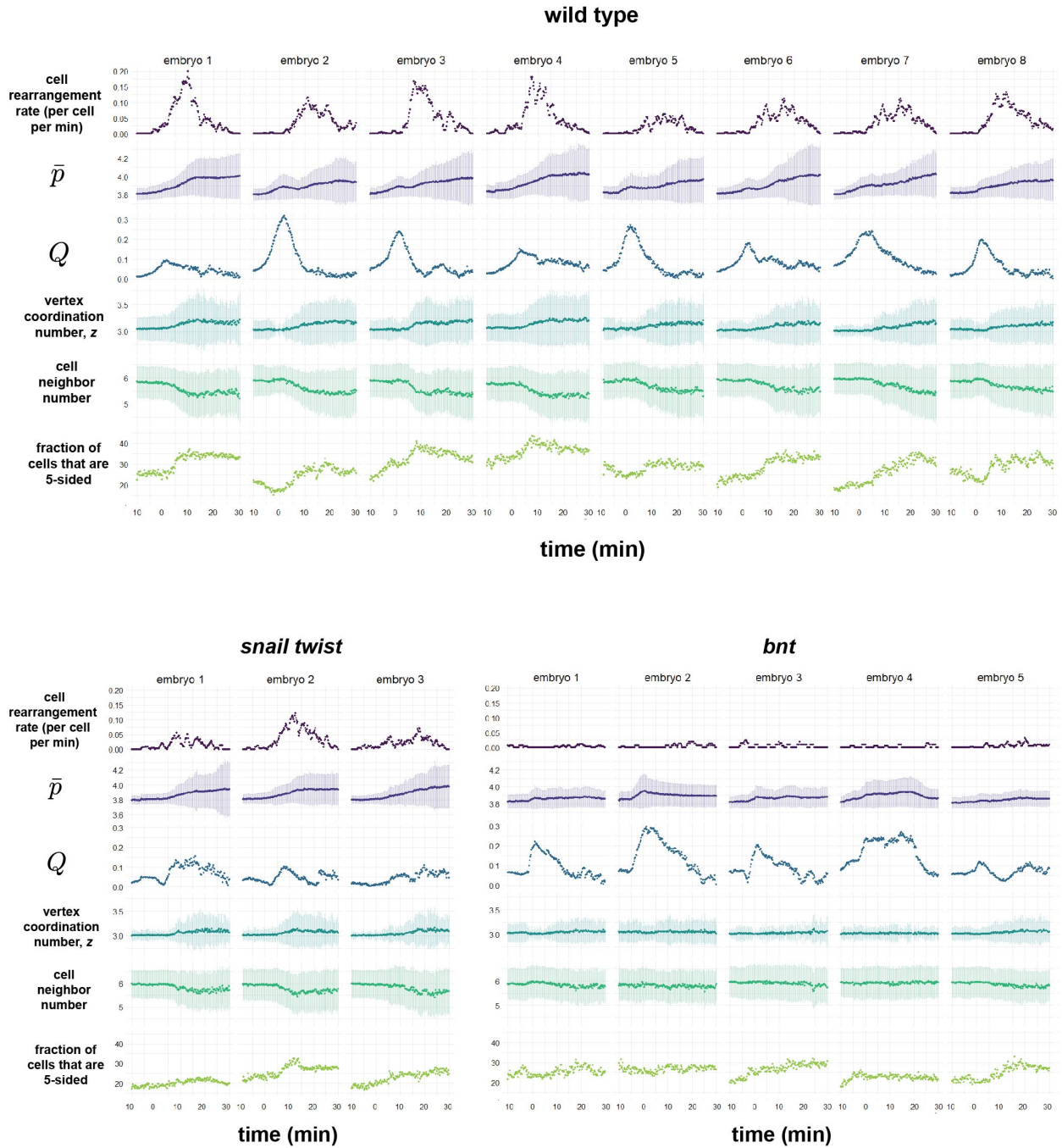

**Figure S2. Behavior of the germband in individual wild-type and mutant embryos over time.** The cell rearrangement rate per cell per minute, average cell shape index $\bar{p}$, cell shape alignment index $Q$, average vertex coordination number $z$, average cell neighbor number, and fraction of cells that are pentagonal $f_5$.

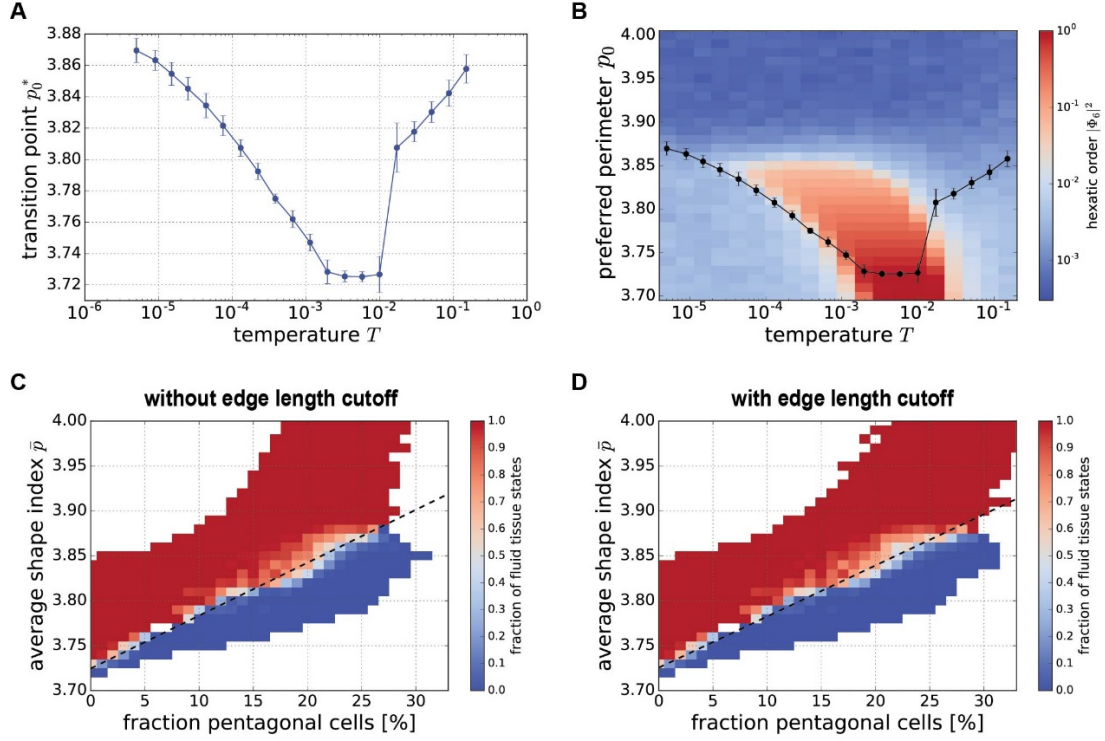

**Figure S3. The vertex model transition point depends on the packing disorder.** Results of vertex model simulations, where the effect of packing disorder is studied by annealing the model tissue with thermal fluctuations prior to quenching to a force-balanced state to create packings with different degrees of disorder. Taken together, these simulation results show that the critical shape index in the vertex model depends on the cellular packing disorder, where fluctuations can help to decrease packing disorder and $p_o^*$. (*A*) The transition point $p_o^*$ decreases with the annealing temperature $T$ before increasing again for very high $T$, confirming a dependence of the transition point on packing disorder. The values for $p_o^*$ decreased from 3.86 for low temperatures to 3.72 for higher temperatures, and increased again for even higher temperatures. (*B*) While the lower bound is below the previously reported value of 3.81, this may be related to partial crystallization of the tissue in this regime. The hexatic bond-orientational order parameter, shown here depending on the preferred shape index $p_0$ and annealing temperature $T$, indicates at least partial crystallization for intermediate temperatures, which correlates with lower transition points $p_o^*$. (*C*) We bin the simulations from panels A, B (100 simulations for each different combination $p_0, T$) with respect to average cell shape index $\bar{p}$ and fraction of pentagonal cells $f_5$. Within each bin we then compute the fraction of fluid states using a cutoff of $10^{-7}$ on the shear modulus. The black dashed line is a linear fit obtained by minimizing the number of simulations on the wrong side of the transition line: $p_0^* = 3.725 + 0.59 f_5$. White regions do not contain any simulation. Same plot as Fig. 2*D* in the main text. (*D*) Same plot as in panel C with a corrected fraction of pentagonal cells $f_5'$. As the spatial resolution in our experiments is limited, we detect cell-cell interfaces with a length smaller than $0.11d$ as manyfold vertices, where $d$ is the square root of the average cell area. We applied the same cutoff to interpret the energy-minimized configurations in panel C, which generally leads to a somewhat higher fraction of pentagonal cells $f_5' > f_5$. However, the linear fit of the solid-fluid transition (black dashed line) is only slightly altered to: $p_0^* = 3.726 + 0.57 f_5'$.

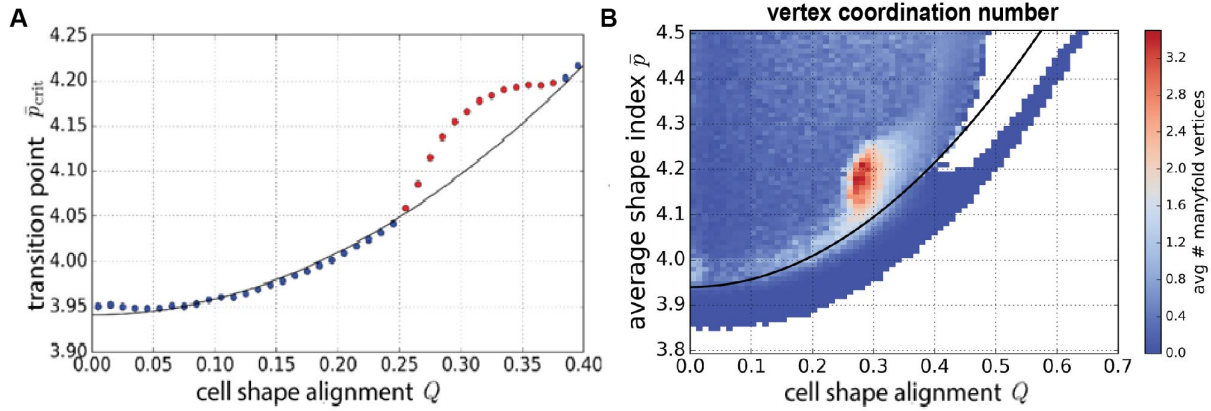

**Figure S4. Fits of Eq. (1) in main text to simulations.** (*A*) Fit of Eq. (1) (black line) to the vertex model simulation results for the case of external deformation (dots, cf. Fig. 3*D*). The average transition points $\bar{p}_{\mathrm{crit}}$ were extracted from the simulation data (Fig. 3*D*) by interpreting the fraction of fluid networks for fixed $Q$ as a cumulative probability density and extracting the average from it (9). From the fit we find $p_o^* = 3.94$ and $b = 0.43$. The red data points were excluded from the fit. These points are related to an excess number of rigid states in the region (cf. panel B and Fig. 3*D*). (*B*) The excess number of rigid states around $Q \approx 0.3$ and $\bar{p} \approx 4.15$ (cf. panel A and Fig. 3*D*) is associated with an increase in the number of manyfold coordinated vertices in model tissues. Plotted here is the total number of manyfold vertices in configurations of 512 cells.

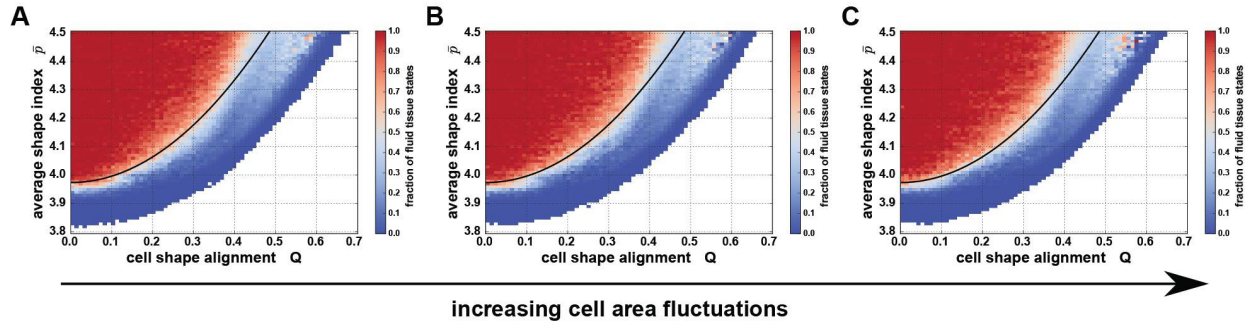

increasing cell area fluctuations

**Figure S5. Effects of cell area variation in vertex model simulations.** Effects of cell area variation on vertex model simulation results. A Gaussian distribution for the preferred cell areas was used with relative standard deviations of 0% (*A*), 10% (*B*), and 20% (*C*). Solid black lines represent fits to the quadratic relation Eq. (1) in the main text. Variation in cell area has only a very small effect on our theoretical findings and the parameters $p_0^*$ and $b$. We find the following fit parameters: (A) $p_0^* = 3.96$, $b = 0.57$; (B) $p_0^* = 3.97$, $b = 0.58$; (C) $p_0^* = 3.97$, $b = 0.57$. These parameters differ somewhat from what we find in Figs. 3*D*, S4*A*, because there we allowed for manyfold vertices as intermediate states during the energy minimization, whereas in these simulations, we did not allow manyfold vertices in order to reduce simulation run time.

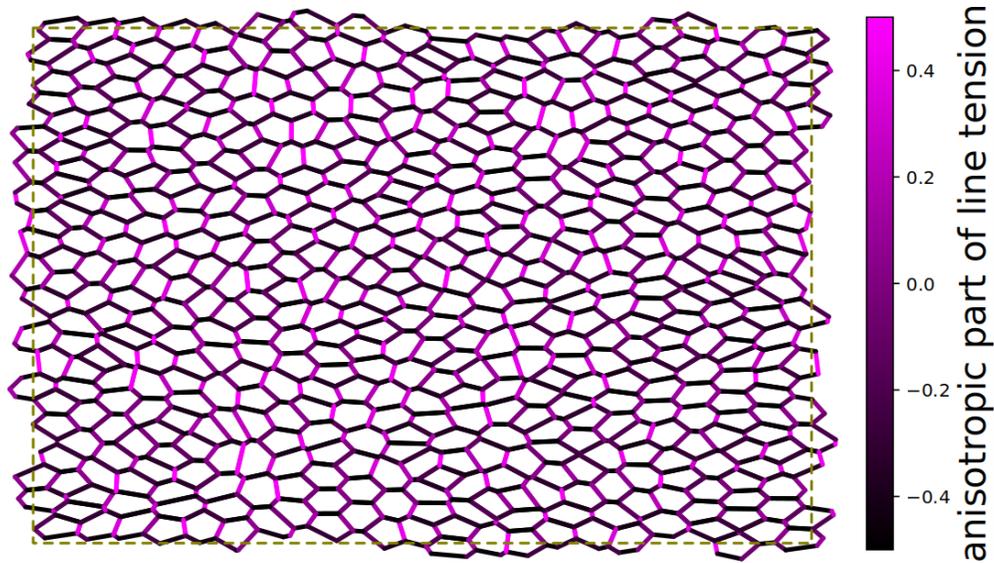

**Figure S6. Vertex model tissue with anisotropic internal stresses.** Force-balanced vertex model state with anisotropic cell-cell interfacial tensions to model the effects of planar polarized myosin II localization patterns. Cell edges color-coded by the tension anisotropy. Parameters: $p_0 = 3.5$, $\lambda_0 = 0.5$.

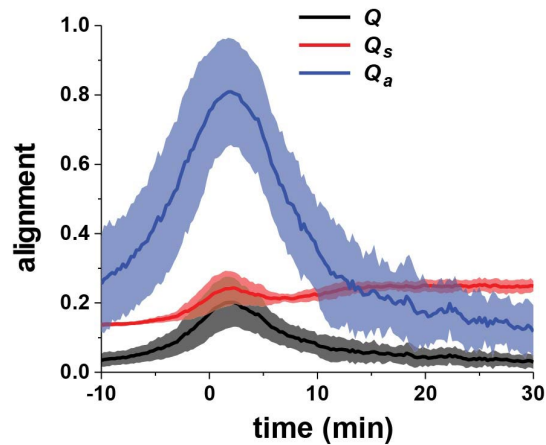

**Figure S7. Cell shape alignment.** Throughout the main text, we discuss the parameter $Q$ (black curve), which describes both cellular shape anisotropy and alignment of cell shapes in the tissue. Indeed, $Q$ can be decomposed as $Q = Q_s Q_a$ (cf. Eq. (S8)), where $Q_s$ represents cellular shape anisotropy (red curve) and $Q_a$ represents cell shape alignment (blue curve). The "pure" nematic alignment parameter $Q_a$ ranges from zero (random cellular orientation) to one (all cells aligned along the same axis, although with potentially different magnitudes). Here we plot $Q$, $Q_s$, and $Q_a$ for the wild-type germband. $Q_a$ is almost one at around $t = 0$, which indicates almost perfect alignment of the cells at that time point.

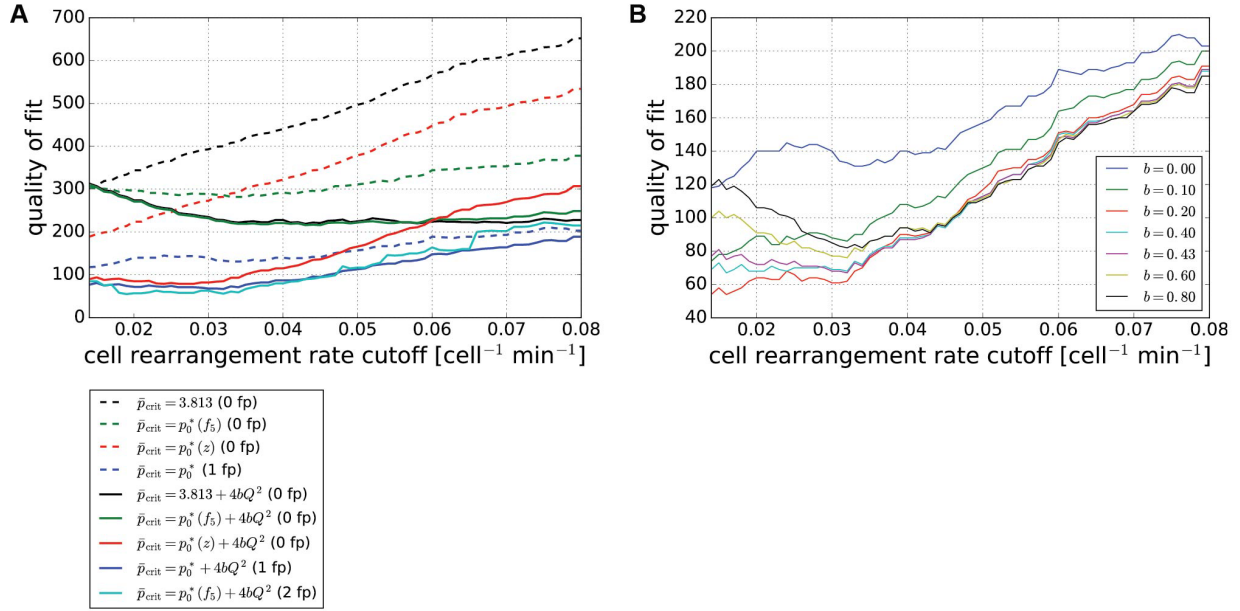

**Figure S8. Quality of fits.** To compare theoretical predictions to experimental data, we define a quality of fit measure $n_{\text{tot}}$, which is the number of experimental data points that are wrongly categorized as either solid or fluid by the prediction. Thus, a better fit is associated with a smaller value of $n_{\text{tot}}$. (*A*) Comparison of several theoretical predictions by plotting the quality of fit, $n_{\text{tot}}$, over a range of cell rearrangement rate cutoffs. The predictions include: a constant value of $p_0^* = 3.813$ (black dashed line) (11), a constant value of $p_0^* = 3.813$ and $Q$-dependent correction according to Eq. (1) in the main text (black solid line), fraction-of-pentagon-dependent $p_0^*(f_5)$ extracted from vertex model simulations in Fig. 2*D* (green dashed line), same with $Q$-dependent correction (green solid line), vertex-coordination-dependent $p_0^*(z)$ according to Ref. (8) (red dashed line), and same with $Q$-dependent correction (red solid line; cf. Eq. (2) and Fig. 4*D*). Note that all of these six predictions are parameter free ("fp" in the legend indicates the respective number of fit parameters used). In addition, we plot the quality of one-parameter fits, where we either fit a constant $p_0^*$ (blue dashed line; cf. Fig. 4*C*) or additionally include the $Q$-dependent correction (blue solid line; cf. Fig. 4*C*). Finally, we plot the quality of a two-parameter fit, where for each cell rearrangement cutoff we extract the fraction-of-pentagon dependence $p_0^*(f_5)$ from experimental data using a fit like Fig. 4*E* and include the $Q$-dependent correction of Eq. (1) (cyan solid line; cf. *SI Appendix* Fig. S9). We find that including the $Q$-dependent correction always improves the prediction, where each time we have fixed $b = 0.43$. Moreover, the best parameter-free prediction for small rearrangement rate cutoffs is given by the red solid line corresponding to Eq. (2) in the main text. (*B*) Quality of fit, $n_{\text{tot}}$, for different fixed values of the parameter $b$ for one-parameter fits of Eq. (1) where we vary $p_0^*$ (cf. blue dashed and solid lines in panel A and Fig. 4*C*). Throughout the manuscript, we used the value of $b = 0.43$ determined from vertex model simulations results. The quality of fit is only slightly improved by changing the value of $b$.

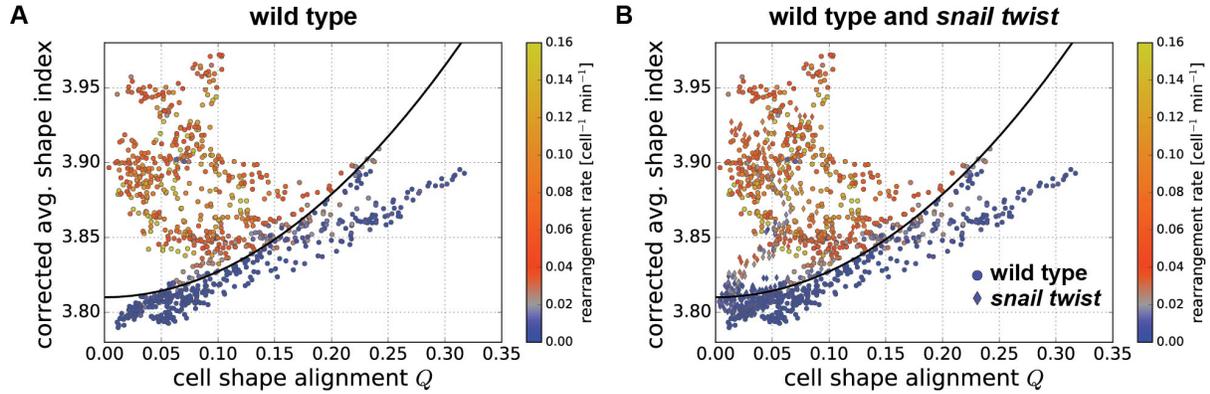

**Figure S9. Accounting for packing disorder using the fraction of pentagonal cells.** In the main text, we account for the effects of packing disorder on $p_o^*$ by taking into account the vertex coordination number in the tissue (cf. Fig. 4*F*) using the previous prediction in Ref. (8), as shown in Eq. (2) in the main text. Alternately, one could use the fraction of pentagonal cells, which in our experiments correlates well with $p_o^*$ (cf. Fig. 4*E*). Here we show the results if we instead correct the cell shape index by the fraction of pentagonal cells at the transition point, according to the linear fit in Fig. 4*E*. (*A*) Same experimental data as in Fig. 4*D*, but with the cell shape corrected by fraction of pentagonal cells observed in the tissue at the transition point. (*B*) Same experimental data as in Fig. 5*H*, but with the cell shape corrected by fraction of pentagonal cells observed in the tissue at the transition point.

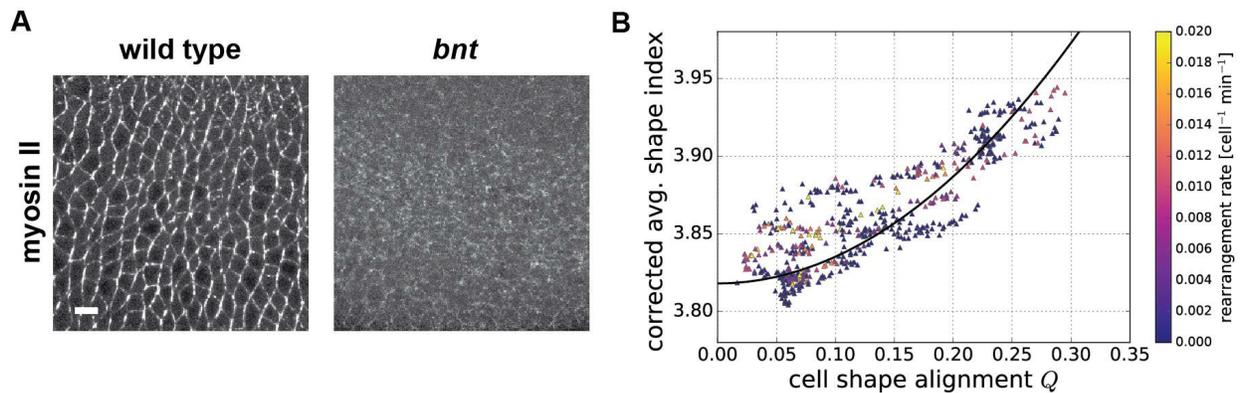

**Figure S10. *bnt* mutants.** (*A*) *bcd nos tsl* (*bnt*) mutant embryos, which lack anterior-posterior patterning genes required for axis elongation, show severely disrupted myosin planar polarity compared to wild-type embryos. Scale bar, 10 μm. (*B*) Same data as in Fig. 5*I*, but with a different color scale to better distinguish rearrangement rate values within *bnt* embryos. Relationship between the corrected average cell shape index $\bar{p}_{corr}$ and $Q$ for 5 *bnt* embryos with each point representing a time point in a single embryo. Instantaneous cell rearrangement rate is represented by the color of each point. Solid line represents the prediction of Eq. (2) in the main text.